\documentclass[12pt]{article}

\usepackage{amsfonts,amssymb,bm,amsmath}
\usepackage{cite}
\usepackage[dvips]{graphicx}

\newcommand{\cg}[3]{C^{#1}_{#2\;#3}}
\newcommand {\beq}{\begin{equation}}
\newcommand {\eeq}{\end{equation}}
\newcommand {\beqa}{\begin{eqnarray}}
\newcommand {\eeqa}{\end{eqnarray}}
\newcommand {\n}{\nonumber \\}
\newcommand {\tr}{\mbox{tr}}
\newcommand {\Tr}{\mbox{Tr}}

\newcommand{\sixj}[6]{{\scriptsize
			\begin{Bmatrix}
			#1 &\hspace{-2mm} #2 &\hspace{-2mm} #3 \\
			#4 &\hspace{-2mm} #5 &\hspace{-2mm} #6
			\end{Bmatrix}
			}}
\newcommand{\ninej}[9]{{\scriptsize
			\begin{Bmatrix}
			#1 &\hspace{-2mm} #2 &\hspace{-2mm} #3 \\
			#4 &\hspace{-2mm} #5 &\hspace{-2mm} #6 \\
			#7 &\hspace{-2mm} #8 &\hspace{-2mm} #9
			\end{Bmatrix}
			}}

\newcommand{\cD}{{\cal D}}

\newcommand{\cF}{{\cal F}}
\newcommand{\cG}{{\cal G}}

\allowdisplaybreaks

\begin{document}
\setlength{\oddsidemargin}{0cm}
\setlength{\baselineskip}{7mm}

\begin{titlepage}
\renewcommand{\thefootnote}{\fnsymbol{footnote}}
\begin{normalsize}
\begin{flushright}
\begin{tabular}{l}
KUNS-2347 \\
June 2011
\end{tabular}
\end{flushright}
\end{normalsize}

~~\\

\vspace*{0cm}
    \begin{Large}
       \begin{center}
          {A non-perturbative formulation of ${\cal N}=4$ super Yang-Mills theory
           based on the large-$N$ reduction}
       \end{center}
    \end{Large}
\vspace{0.5cm}

\begin{center}
Goro Ishiki$^{1),2)}$\footnote
            {
e-mail address : 
ishiki@post.kek.jp},
Shinji Shimasaki$^{2),3)}$\footnote
            {
e-mail address : 
shinji@gauge.scphys.kyoto-u.ac.jp}
        and
Asato Tsuchiya$^{4)}$\footnote
            {
e-mail address : satsuch@ipc.shizuoka.ac.jp}\\
      \vspace{0.5cm}
                    
       $^{1)}$ {\it Center for Quantum Spacetime (CQUeST)
                    }\\
               {\it Sogang University, Seoul 121-742, Republic of Korea}\\
      \vspace{0.3cm}
      $^{2)}$ {\it Department of Physics, Kyoto University, Kyoto 606-8502, Japan}\\
      \vspace{0.3cm}
      $^{3)}$ {\it Harish-Chandra Research Institute}\\
               {\it Chhatnag Road, Jhusi,
                Allahabad 211019, India }\\
      \vspace{0.3cm}
      $^{4)}$ {\it Department of Physics, Shizuoka University}\\
               {\it 836 Ohya, Suruga-ku, Shizuoka 422-8529, Japan}
               
\end{center}

\vspace{1cm}

\begin{abstract}
\noindent
We study a non-perturbative formulation of ${\cal N}=4$ super Yang-Mills theory (SYM)
on $R\times S^3$ proposed in arXiv:0807.2352.
This formulation is based on the large-$N$ reduction, and 
the theory can be described
as a particular large-$N$ limit of 
the plane wave matrix model (PWMM), which is obtained 
by dimensionally reducing the original theory over $S^3$. 
In this paper, we perform some tests for this proposal.
We construct an operator in the PWMM
that corresponds to the Wilson loop in SYM in the continuum limit
and calculate the vacuum expectation value 
of the operator for the case of the circular contour.
We find that our result indeed agrees with 
the well-known result first obtained by Erickson, Semenoff and Zarembo.
We also compute the beta function at the 1-loop level based on 
this formulation and see that it is indeed vanishing.
\end{abstract}
\vfill
\end{titlepage}
\vfil\eject

\setcounter{footnote}{0}

\section{Introduction}
\setcounter{equation}{0}
The AdS/CFT correspondence \cite{Maldacena:1997re,Gubser:1998bc,Witten:1998qj} 
has been intensively studied and fruitfully extended
to various directions over a decade. However, a complete proof of this conjecture
is still missing even for the most typical example, 
the correspondence between ${\cal N}=4$ super Yang-Mills theory (SYM)
and type IIB superstring theory on $AdS_5\times S^5$.
This is partially because the correspondence is a strong/weak one.
Namely, the region on the string theory side in which the supergravity approximation 
or the classical string approximation is valid is mapped to the strongly coupled region
in the planar limit on the gauge theory side.
In order to study ${\cal N}=4$ SYM in the strongly coupled region,
one needs to have a non-perturbative formulation such as the lattice gauge theory. 
In fact, there are considerable developments in the lattice theories of ${\cal N}=4$ SYM 
\cite{Kaplan:2005ta,Unsal:2005us,Elliott:2008jp,Catterall:2008dv,Giedt:2009yd,Hanada:2010kt,Hanada:2010gs,
Catterall:2011pd}.
However, any lattice formulations of ${\cal N}=4$ SYM proposed so far seem to require
the fine-tuning of at least three parameters\footnote{It was shown recently that only a fine-tuning of one parameter
is needed at least to one-loop order in a lattice formulation \cite{Catterall:2011pd}. It is discussed that
no fine-tuning is required in the recent proposals\cite{Hanada:2010kt,Hanada:2010gs}.}.

In \cite{Ishii:2008ib}, a non-perturbative formulation of ${\cal N}=4$
SYM on $R\times S^3$ in the planar ('t Hooft) limit
was proposed by using the plane wave matrix model (PWMM) \cite{Berenstein:2002jq} 
(for earlier discussions, see \cite{Ishiki:2006yr,Ishii:2008tm}).
Note that ${\cal N}=4$ SYM on $R^4$ at a conformal point has the $PSU(2,2|4)$ 
symmetry with thirty-two supercharges and is equivalent to
${\cal N}=4$ SYM on $R\times S^3$ through the conformal mapping.
The PWMM can be obtained by dimensionally reducing ${\cal N}=4$ SYM on $R\times S^3$ over $S^3$ \cite{Kim:2003rza,Lin}.
In the formulation, ${\cal N}=4$ SYM on $R\times S^3$ is retrieved by taking an appropriate large-$N$ limit for
the theory around a particular vacuum of the PWMM.
Thus the formulation is viewed as
an extension of the large-$N$ reduction \cite{largeNreduction}, 
which asserts that the planar limit of a gauge theory is described by its dimensionally
reduced model.
The formulation provides a matrix regularization of the planar ${\cal N}=4$ SYM on $R\times S^3$, 
where the matrix size plays a role of the ultraviolet cutoff.
Remarkably, the formulation preserves sixteen supercharges and the gauge symmetry.
Note that this number of the preserved supercharges is optimal, because
any regularization must break the conformal symmetry so that the number is
inevitably reduced to less than or equal to sixteen from thirty-two.
Since the formulation preserves so many supercharges and provides a massive theory,
no fine-tuning should be needed in taking the continuum limit. 
This is advantageous to the lattice
formulations of ${\cal N}=4$ SYM, while the latter may also be used to
study the finite $N$ ${\cal N}=4$ SYM.
Thus the formulation gives a feasible way to simulate the planar ${\cal N}=4$ SYM 
by computer using the methods in \cite{Anagnostopoulos:2007fw,Catterall:2008yz} 
and study its strong coupling dynamics\footnote{Monte Carlo simulation of
the correlation functions of the chiral primary
operators in ${\cal N}=4$ SYM and comparison of the results with the prediction from the gravity side will
be reported in \cite{HIKNT}. Preliminary 
results including the one for the Wilson loop are seen in \cite{Nishimura:2009xm,Honda:2010nx}. For related work,
see \cite{Berenstein:2007wz,Berenstein:2008jn,Catterall:2010gf}.}.

To check the validity of the formulation, some one-loop perturbative calculations were performed in \cite{Ishii:2008ib}.
It was shown that the tadpole vanishes and that the fermion one-loop self-energy agrees with 
the one in the continuum theory. 
In \cite{Ishiki:2008te,Ishiki:2009sg}, the known result of the confinement-deconfinement transition 
in the planar ${\cal N}=4$ SYM on $R\times S^3$
in the weak coupling limit at finite temperature was reproduced in the formulation.
In \cite{Kitazawa:2008mx}, a two-loop calculation in the high temperature limit was done, and
the result was consistent with the continuum theory.  
In \cite{Ishiki:2009vr,Ishiki:2010pe}, based on the earlier work \cite{Ishii:2007sy,Ishiki:2008vf},
the same large-$N$ reduction for $S^3$ was applied to Chern-Simons theory
on $S^3$ to obtain a matrix regularization of the theory in the planar limit, and
the exact results on Chern-Simons theory on $S^3$ were reproduced.
Extension of the formulation to other supersymmetric gauge theories 
was discussed in \cite{Hanada:2009kz,Hanada:2009hd}. 
In \cite{Kawai:2009vb,Kawai:2010sf}, large-$N$ reduction on general group manifolds and coset spaces
was found. Applying the case of $SU(2)\simeq S^3$ to planar ${\cal N}=4$ SYM on $R\times S^3$ yields
another non-perturbative formulation of the theory in terms of the PWMM.
Recently, a matrix model regularization of ${\cal N}=4$ SYM on $S^4$ was proposed 
in \cite{Heckman:2011ju}.

In this paper, we make a further check by calculating 
two quantities: the vacuum expectation value (VEV)
of a half-BPS Wilson loop and the beta function.

Erickson, Semenoff and Zarembo gave 
a successful example of calculation in the strongly coupled regime of ${\cal N}=4$ SYM \cite{Erickson:2000af}
(see also \cite{Drukker:2000rr}).
They considered (locally) supersymmetric Wilson loops in ${\cal N}=4$ SYM on $R^4$.
It is conjectured that the logarithms of 
the VEV of such a Wilson loop in ${\cal N}=4$ SYM corresponds
to minus the area of the minimal surface of a string world sheet
in the $AdS_5\times S^5$ such that the boundary of the world sheet 
coincides with the loop \cite{Rey:1998ik,Maldacena:1998im}.
In the planar limit, they evaluated the VEV 
of the circular Wilson loop, which is half-BPS,
by summing up all ladder diagrams of
all orders in the perturbative expansion.
They indeed showed that in the strong coupling region 
the result agrees with the prediction
on the gravity side \cite{Berenstein:1998ij,Drukker:1999zq}. Thus, they gave a nontrivial check of 
the AdS/CFT correspondence.
Their result was reproduced through the localization method
in \cite{Pestun:2007rz} to show that it is indeed the exact result.
In this paper, we reproduce their result in the above formulation of ${\cal N}=4$ SYM on $R\times S^3$.
We first construct an operator in terms of the matrices
in such a way that it coincides with the Wilson loop in ${\cal N}=4$ SYM
on $R\times S^3$ in the continuum limit.
Then, we concentrate on the circular loop,
and calculate its VEV
by summing up all the planar ladder diagrams. 
We find that the VEV exactly agrees with the result in \cite{Erickson:2000af}.

We also evaluate the 1-loop beta function and find that 
it indeed vanishes, which is consistent with restoration of the superconformal invariance.

This paper is organized as follows. 
In section 2, we review the non-perturbative formulation of the planar
${\cal N}=4$ SYM on $R\times S^3$ proposed in \cite{Ishii:2008ib}.
In section 3, after reviewing Wilson loops in the AdS/CFT correspondence,
we construct operators in the non-perturbative formulation
which correspond to the Wilson loops in ${\cal N}=4$
SYM on $R\times S^3$ in the continuum limit.
In section 4, we calculate the VEV of the operator in the half-BPS case
and show that it agrees with the known exact result.
In section 5, we calculate the 1-loop beta function and verify that
it indeed vanishes.
Section 6 is devoted to summary and discussion. 
In appendices, some details are gathered.

\section{Large-$N$ reduction for ${\cal N}=4$ SYM on $R\times S^3$}
\setcounter{equation}{0}

In this section, we review the novel large-$N$ reduction
for ${\cal N}=4$ SYM on $R\times S^3$ proposed in \cite{Ishii:2008ib}.
In section 2.1, as a warm-up, we review the large-$N$ reduction on $S^1$ developed in \cite{Ishii:2008ib},
by taking the $\phi^3$ theory on $S^1$ as an example.
In section 2.2, to explain the mechanism of the large-$N$ reduction on $S^3$, 
we consider the large $N$ reduction for the $\phi^3$ theory on $S^3$. 
In section 2.3, we apply the mechanism 
to ${\cal N}=4$ SYM on $R\times S^3$.

\subsection{Large-$N$ reduction on $S^1$}
We consider the $\phi^3$ matrix quantum mechanics at finite temperature with
the inverse temperature $R$. In other words, it is a one-dimensional matrix field theory
on $S^1$ with the radius $R$. The action is 
\begin{align}
S=\frac{1}{g^2}\int_0^{2\pi R} dx \; {\rm Tr} \left(\frac{1}{2}\left(\frac{d\phi}{dx}\right)^2+\frac{\xi^2}{2R^2}\phi^2
+\frac{1}{3R}\phi^3 \right) \ ,
\label{phi^3 theory on S^1}
\end{align}
where $0 \leq x <2\pi R$, $\phi$ is an $N\times N$ hermitian matrix, and $\xi$ is a dimensionless mass.

To diagonalize the quadratic part in (\ref{phi^3 theory on S^1}), we make the Fourier expansion
\begin{align}
\phi(x)=\sum_n\phi_ne^{i\frac{n}{R}x} \ .
\label{Fourier expansion}
\end{align}
The action (\ref{phi^3 theory on S^1}) is rewritten in terms of the Fourier modes as
\begin{align}
S=\frac{V_{S^1}}{g^2}\left[\frac{1}{2R^2}\sum_n(n^2+\xi^2)\phi_n\phi_{-n}
+\frac{1}{3R}\sum_{n_1,n_2,n_3}\delta_{n_1+n_2+n_3,0}\phi_{n_1}\phi_{n_2}\phi_{n_3} \right] \ ,
\end{align}
where $V_{S^1}=2\pi R$ is the volume of $S^1$.

Here, for instance, we consider the following correlation function
\begin{align}
\left\langle \frac{1}{N}\mbox{Tr}(\phi(x_1)\phi(x_2))\right\rangle \ .
\label{correlation function on S^1}
\end{align}
By using the translation invariance of the theory and the Fourier expansion (\ref{Fourier expansion}), 
we see that the above correlation function is equal to
\begin{align}
\frac{1}{2\pi R}\int dy \left\langle\frac{1}{N}\mbox{Tr}(\phi(x_1-x_2+y)\phi(y))\right\rangle
=\sum_n\left\langle\frac{1}{N}\mbox{Tr}(\phi_n\phi_{-n})\right\rangle e^{i\frac{n}{R}(x_1-x_2)} \ .
\label{correlation function on S^1 2}
\end{align}
We calculate a diagram in Fig. 1 which appears in the perturbative expansion of 
$\langle \frac{1}{N}\mbox{Tr}(\phi_n\phi_{-n})\rangle$:
\begin{align}
\left(\frac{g^2N}{V_{S^1}}\right)^2R^6\left(\frac{1}{n^2+\xi^2}\right)^2
\sum_l \frac{1}{l^2+\xi^2}\frac{1}{(n+l)^2+\xi^2} \ .
\label{correlation function on S^1 3}
\end{align}
This survives in the planar limit where
\begin{align}
N\rightarrow\infty \;\; \mbox{with}\;\; g^2N \;\; \mbox{fixed} \ .
\label{planar limit}
\end{align}
In this limit, all the planar diagrams such as Fig. 1 contribute to (\ref{correlation function on S^1}),
while other non-planar diagrams such as Fig. 2 do not.

\begin{figure}[tbp]
\begin{center}
 \begin{minipage}{0.3\hsize}
  \begin{center}
\includegraphics[width=30mm]{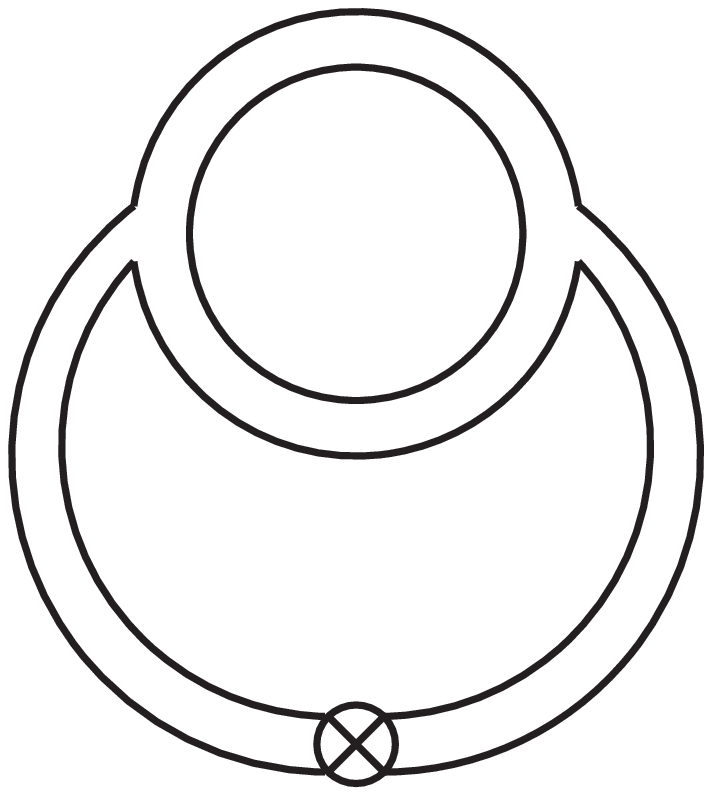}
  \caption{Planar diagram for the two-point correlation function}
  \label{planar}
\end{center}
 \end{minipage}
\hspace*{1cm}
 \begin{minipage}{0.3\hsize}
  \begin{center}
  {\includegraphics[width=30mm]{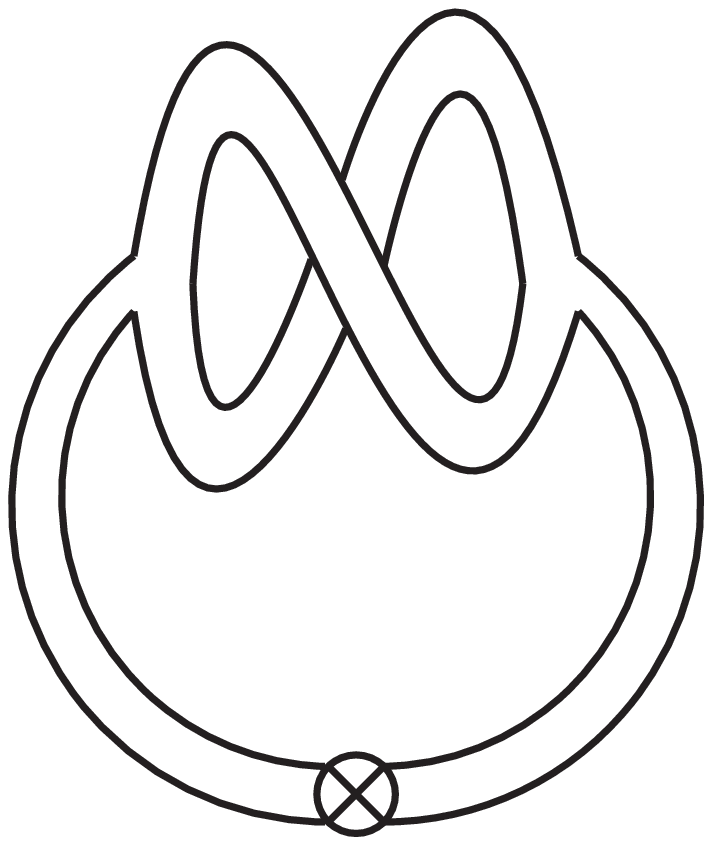}}
  \caption{Non-planar diagram for the two-point correlation function}
  \label{nonplanar}
\end{center}
 \end{minipage}
\end{center}
\label{Diagrams for the two-point correlation function}
\end{figure}

To obtain the reduced model of (\ref{phi^3 theory on S^1}), 
we introduce a constant diagonal matrix
with the eigenvalues uniformly distributed:
\begin{align}
P=\frac{1}{R}\mbox{diag}\left(\frac{-\nu+1}{2},\frac{-\nu+3}{2},\cdots,\frac{\nu-1}{2}\right)
\otimes {\bf 1}_k \ ,
\label{P}
\end{align}
where $\nu$ and $k$ are integers satisfying\footnote{The tensor product of ${\bf 1}_k$ is needed
for the theory on $S^1$ to extract the planar contribution through the $k\rightarrow\infty$ limit \cite{Ishii:2008ib}. 
It is not needed for the theory on $R$.} 
\begin{align}
N=\nu k \ .
\label{N}
\end{align}
$\nu$ turns out to play the role of the UV cutoff.
The rule for obtaining the reduced model is
\begin{align}
\phi(x)\rightarrow e^{iPx}\phi e^{-iPx}, \;\;\; V_{S^1}\rightarrow v \ ,
\label{rule for S^1}
\end{align}
where $\phi$ is an $N\times N$ hermitian matrix independent of $x$. 
$v$ is determined shortly such that the reduced model reproduces the original theory.
The rule (\ref{rule for S^1}) implies that
\begin{align}
\frac{d\phi(x)}{dx}\rightarrow ie^{iPx}[P,\phi]e^{-iPx} \ .
\end{align}
Then, applying the rule (\ref{rule for S^1})  to (\ref{phi^3 theory on S^1}) yields the reduced model
of (\ref{phi^3 theory on S^1})
\begin{align}
S_r=\frac{v}{g^2}\mbox{Tr}\left(-\frac{1}{2}[P,\phi]^2+\frac{\xi^2}{2R^2}\phi^2+\frac{1}{3R}\phi^3\right) \ .
\label{reduced model of phi^3 theory on S^1}
\end{align}
If $\phi$ is decomposed to a tensor product of a $\nu \times \nu$ matrix and
a $k\times k$ matrix following (\ref{P}), (\ref{reduced model of phi^3 theory on S^1}) is expressed as 
\begin{align}
S_r=\frac{v}{g^2}\mbox{tr}\left( \frac{1}{2R^2}\sum_{s,t}((s-t)^2+\xi^2))\phi^{(s,t)}\phi^{(t,s)}
+\frac{1}{3R}\sum_{s,t,u}\phi^{(s,t)}\phi^{(t,u)}\phi^{(u,s)} \right) \ ,
\end{align}
where $\phi^{(s,t)}$ is a $k\times k$ matrix, 
and tr stands for the trace over $k\times k$ matrices.
The range of the indices $s,t,u$ is
\begin{align}
\frac{-\nu+1}{2} \leq s,t,u \leq \frac{\nu-1}{2} \ ,
\end{align}
and the indices $s,t,u$ run integers for odd $\nu$ and half-integers for even $\nu$ .

By applying the rule (\ref{rule for S^1}), we obtain the observable corresponding to 
(\ref{correlation function on S^1})
\begin{align}
\left\langle \frac{1}{N}\mbox{Tr}(e^{iPx_1}\phi e^{-iPx_1}e^{iPx_2}\phi e^{-iPx_2}) \right\rangle_r \ ,
\label{correlation function in reduced model of phi^3 theory on S^1}
\end{align}
where $\langle \cdots \rangle_{r}$ stands for the VEV in the reduced model.
By using (\ref{P}), this is rewritten as
\begin{align}
\frac{1}{\nu}\sum_{s,t}\left\langle \frac{1}{k}\mbox{tr}(\phi^{(s,t)}\phi^{(t,s)})\right\rangle_r
e^{i\frac{s-t}{R}(x_1-x_2)} \ .
\label{correlation function in reduced model of phi^3 theory on S^1 2}
\end{align}
We calculate again the diagram in Fig. 1 which appears in the perturbative expansion
of $\langle \frac{1}{k}\mbox{tr}(\phi^{(s,t)}\phi^{(t,s)})\rangle_r$:
\begin{align}
\sum_{u}\left(\frac{g^2N}{\nu v}\right)^2R^6 \left(\frac{1}{(s-t)^2+\xi^2}\right)^2
\frac{1}{(t-u)^2+\xi^2}\frac{1}{(u-s)^2+\xi^2} \ .
\label{correlation function of reduced model of phi^3 theory on S^1 3}
\end{align}

Comparing (\ref{correlation function on S^1 3}) and 
(\ref{correlation function of reduced model of phi^3 theory on S^1 3}),
we put
\begin{align}
v=\frac{V_{S^1}}{\nu} \ ,
\label{v in reduced model of phi^3 theory on S^1}
\end{align}
and take a limit
\begin{align}
k\rightarrow\infty, \;\; \nu\rightarrow\infty \;\;\mbox{with}\;\; g^2N \;\;\mbox{fixed} \ .
\label{limit in reduced model}
\end{align}
Note that (\ref{limit in reduced model}) implies (\ref{planar limit}) because $N=\nu k$.
We make an identification $n=s-t$, $l=t-u$ and $-n-l=u-s$ , and find that 
(\ref{correlation function of reduced model of phi^3 theory on S^1 3}) agrees 
with (\ref{correlation function on S^1 3}) in the limit 
(\ref{limit in reduced model}).
This correspondence holds for all the planar diagrams such as Fig. 1. As for the non-planar diagrams
such as Fig. 2, such correspondence does not hold. However, because of $k\rightarrow\infty$,
only the planar diagrams contribute to (\ref{correlation function in reduced model of phi^3 theory on S^1 2}).
Furthermore, we compare (\ref{correlation function in reduced model of phi^3 theory on S^1 2}) with
(\ref{correlation function on S^1 2}). While the summation over $s$ and $t$ in 
(\ref{correlation function in reduced model of phi^3 theory on S^1 2})
is redundant compared to the summation over $n$ in (\ref{correlation function on S^1 2}), this redundancy
is canceled by a factor $1/\nu$ in (\ref{correlation function in reduced model of phi^3 theory on S^1 2}).
Thus we conclude that (\ref{correlation function in reduced model of phi^3 theory on S^1}) in
the limit (\ref{limit in reduced model}) agrees with (\ref{correlation function on S^1}) in the 
planar limit.

In a similar manner, one can easily show that in the limit (\ref{limit in reduced model})
\begin{align}
\left\langle\frac{1}{N}\mbox{Tr}(e^{iPx_1}\phi e^{-iPx_1} e^{iPx_2}\phi e^{-iPx_2} \cdots e^{iPx_r}\phi e^{-iPx_r})
\right\rangle_r
=\left\langle\frac{1}{N}\mbox{Tr}(\phi(x_1)\phi(x_2)\cdots\phi(x_r))\right\rangle \ .
\label{correspondence of general correlation functions}
\end{align}
One can also show the following equality for the free energies of both theories 
in the limit (\ref{limit in reduced model}):
\begin{align}
\frac{F_r}{N^2v}=\frac{F}{N^2V_{S^1}} \ .
\label{correspondence of free energies on S^1}
\end{align}
This gives an interpretation of $v$ that it is the volume of space on which the reduced model is defined.
Thus, in the limit (\ref{limit in reduced model}), the reduced model (\ref{reduced model of phi^3 theory on S^1})
with (\ref{v in reduced model of phi^3 theory on S^1}) retrieves 
the planar limit of the original theory (\ref{phi^3 theory on S^1}).

\subsection{Large-$N$ reduction on $S^3$}
To illustrate the large-$N$ reduction on $S^3$, we consider the $\phi^3$ matrix field theory on $S^3$:
\begin{align}
S=\frac{V_{S^3}}{g^2}\int \frac{d\Omega_3}{2\pi^2}
\mbox{Tr}\left(-\frac{\mu^2}{2}({\cal L}_i\phi)^2+\frac{\xi^2\mu^2}{2}\phi^2+\frac{\mu}{3}\phi^3\right).
\label{phi^3 theory}
\end{align}
Here the radius of the $S^3$ is $2/\mu$, and its volume $V_{S^3}$ is given by
\begin{align}
V_{S^3}=\frac{16\pi^2}{\mu^3} \ .
\end{align}
$d\Omega_3$ represents the volume element of unit $S^3$.
$\phi(\Omega_3)$ is an $N\times N$ hermitian matrix scalar field on $S^3$.
${\cal L}_i$ is the Killing vector on unit $S^3$.
Note that $S^3$ is identified with the $SU(2)$ group manifold.
From this point of view, $d\Omega_3$ is the Haar measure which is left and right invariant, and
${\cal L}_i$ is the generator of the left translation on the group manifold.
For an explicit form of $d\Omega_3$ and ${\cal L}_i$, see (\ref{Haar measure}) and (\ref{Killing vector}),
respectively.
$\xi$ is a dimensionless mass as before.

In what follows, the argument goes parallel to the one in the previous subsection.
Indeed, $S^3$ is an $S^1$-bundle over $S^2$, and
this $S^1$ is a counterpart of the $S^1$ on which the $\phi^3$ theory is defined in the previous subsection.
 
To diagonalize the quadratic part in (\ref{phi^3 theory}),
we expand $\phi$ in terms of the spherical harmonics on $S^3$ defined in (\ref{spherical harmonics on S^3}):
\begin{align}
\phi(\Omega_3)=\sum_{J}\sum_{m,\tilde{m}=-J}^J\phi_{Jm\tilde{m}}Y_{Jm\tilde{m}}(\Omega_3),
\label{harmonic expansion}
\end{align}
where $J$ take non-negative integers and half-integers. The properties of the spherical
harmonics on $S^3$ are summarized in appendix A. 
$\tilde{m}$ is viewed as the momentum along the $S^1$ fiber \cite{Ishii:2008ib,Ishiki:2006yr,Ishii:2008tm}.
By using them, the action (\ref{phi^3 theory}) is rewritten
in terms of the modes as
\begin{align}
S=&\frac{V_{S^3}}{g^2}\left[ \frac{\mu^2}{2}
\sum_{Jm\tilde{m}}(-1)^{m-\tilde{m}}(J(J+1)+\xi^2)\mbox{Tr}(\phi_{Jm\tilde{m}}\phi_{J-m-\tilde{m}}) \right.\nonumber\\
&\qquad\left.+\frac{\mu}{3}\sum_{J_1m_1\tilde{m}_1J_2m_2\tilde{m}_2J_3m_3\tilde{m}_3}
{\cal C}_{J_1m_1\tilde{m}_1\:J_2m_2\tilde{m}_2\:J_3m_3\tilde{m}_3}
\mbox{Tr}(\phi_{J_1m_1\tilde{m}_1}\phi_{J_2m_2\tilde{m}_2}\phi_{J_3m_3\tilde{m}_3})\right],
\end{align}
where ${\cal C}_{J_1m_1\tilde{m}_1\:J_2m_2\tilde{m}_2\:J_3m_3\tilde{m}_3}$ is defined in (\ref{calC}). 

By identifying the $S^3$ with the $SU(2)$ group manifold, we denote $\phi(\Omega_3)$ by $\phi(g)$, 
where $g \in SU(2)$. We consider the following correlation function which is a counterpart
of (\ref{correlation function on S^1}):
\begin{align}
\left\langle \frac{1}{N}\mbox{Tr}(\phi(g_1)\phi(g_2))\right\rangle.
\label{correlation function on S^3}
\end{align}
The invariance of the theory under the right translation implies that
(\ref{correlation function on S^3}) is equivalent to
\begin{align}
\int \frac{d\Omega_3}{2\pi^2} \left\langle \frac{1}{N}\mbox{Tr}(\phi(g_1g_2^{-1}g)\phi(g))\right\rangle.
\end{align}
We substitute (\ref{harmonic expansion}) into the above quantity:
\begin{align}
&\sum_{Jm\tilde{m}J'm'\tilde{m}'}\int \frac{d\Omega_3}{2\pi^2}
\left\langle\frac{1}{N}\mbox{Tr}(\phi_{Jm\tilde{m}}\phi_{J'm'\tilde{m}'})\right\rangle
R_J(g_2g_1^{-1})_{m''m}Y_{Jm''\tilde{m}}(\Omega_3)Y_{J'm'\tilde{m}'}(\Omega_3) \nonumber\\
&=\sum_{Jm\tilde{m}m'}(-1)^{m'+\tilde{m}}R_J(g_2g_1^{-1})_{-m'm}
\left\langle\frac{1}{N}\mbox{Tr}(\phi_{Jm\tilde{m}}\phi_{Jm'-\tilde{m}})\right\rangle,
\label{calculation of correlation function in phi^3 theory}
\end{align}
where $R_J(g)$ is the representation matrix of $g$ in the spin $J$ representation, and we have used
(\ref{spherical harmonics on S^3}) and  (\ref{orthonormality}).
We have obtained a counterpart of (\ref{correlation function on S^1 2}).
We calculate a diagram in Fig.\ref{planar} which appears in the perturbative expansion of 
$\langle\frac{1}{N}\mbox{Tr}(\phi_{Jm\tilde{m}}\phi_{Jm'-\tilde{m}})\rangle$:
\begin{align}
\sum_{Jmm'\tilde{m}J_2m_2\tilde{m}_2J_3m_3\tilde{m}_3}
&\frac{(-1)^{m+m'}}{\mu^6}\left(\frac{g^2N}{V_{S^3}}\right)^2
\left(\frac{1}{J(J+1)+\xi^2}\right)^2
\frac{(-1)^{m_2-\tilde{m}_2}}{J_2(J_2+1)+\xi^2}\frac{(-1)^{m_3-\tilde{m}_3}}{J_3(J_3+1)+\xi^2} \nonumber\\
&\times{\cal C}_{J-m-\tilde{m}\:J_2m_2\tilde{m}_2\:J_3m_3\tilde{m}_3}
{\cal C}_{J-m'\tilde{m}\:J_3-m_3-\tilde{m}_3\:J_2m_2-\tilde{m}_2}.
\label{diagram in phi^3 theory}
\end{align}
This survives in the planar limit (\ref{planar limit}). As before,
in this limit, all the planar diagrams such as Fig.\ref{planar} contribute 
to (\ref{correlation function on S^3}), while
other non-planar diagrams such as Fig.\ref{nonplanar} do not.

Next we construct the reduced model of (\ref{phi^3 theory}).
For this it is convenient to introduce an $N$-dimensional
reducible representation of $SU(2)$ in which the generators take the form
\begin{align}
L_i=\bigoplus_{s=-(\nu-1)/2}^{(\nu-1)/2}
\Bigl( L_i^{[j_s]}\otimes {\bf 1}_{k_s} \Bigr) \  ,
\label{reducible representation}
\end{align}
which represents multiple fuzzy spheres.
Here $L_i^{[j]}$ are
the spin $j$ representation matrices of
the $SU(2)$ generators obeying
\begin{align}
[L_i^{[j]},L_j^{[j]}]=i \, \epsilon_{ijk} \, L_k^{[j]}.
\end{align}
$\nu$ is the number of different irreducible representations of $SU(2)$, and
the index $s$ runs integers for odd $\nu$ and half-integers for even $\nu$.
The parameters $j_s$ and $k_s$ in (\ref{reducible representation})
satisfy the relation 
\begin{align}
\sum_{s=-(\nu-1)/2}^{(\nu-1)/2}(2j_s+1)k_s=N .
\end{align}
In particular, in order to construct a reduced model of (\ref{phi^3 theory}), 
we choose the following parameters\footnote{The correspondence between the notation here and 
that in \cite{Ishii:2008ib} is 
$(n)_{\mbox{here}}=(N_0)_{\mbox{there}}$, 
$(\nu)_{\mbox{here}}=(T+1)_{\mbox{there}}$, 
$(k_s)_{\mbox{here}}=(N_s)_{\mbox{there}}$,
$(k)_{\mbox{here}}=(N)_{\mbox{there}}$.}:
\begin{align}
2j_s+1=n+s,\;\; k_s=k.
\label{parameters}
\end{align}
$n$ is the centered value of the dimension of the representations, so that
\begin{align}
N=n\nu k.
\end{align}
The rule for obtaining the reduced model, which is a counterpart of (\ref{rule for S^1}), is
\begin{align}
\phi(g) \rightarrow G^{-1}\phi G, \;\;\; V_{S^3} \rightarrow v,
\label{rule}
\end{align}
where $G$ is the representation matrix of $g$ in the reducible representation specified
by (\ref{parameters}), and $\phi$ is an $N\times N$ hermitian matrix independent of $\Omega_3$.
$v$ is determined shortly such that the reduced model reproduces the original theory.
By using (\ref{left translation}), it is easy to see that
\begin{align}
{\cal L}_i\phi \rightarrow G^{-1}[L_i,\phi]G.
\end{align}
Then applying the rule to (\ref{phi^3 theory}) yields the reduced model of (\ref{phi^3 theory})
\begin{align}
S_r=\frac{v}{g^2}\mbox{Tr}\left(-\frac{\mu^2}{2}[L_i,\phi]^2+\frac{\xi^2\mu^2}{2}\phi^2
+\frac{\mu}{3}\phi^3\right).
\label{reduced model of phi^3 theory}
\end{align}

To see how this reduced model retrieves the original theory, 
we expand the $(s,t)$ block of $\phi$ we denote by $\phi^{(s,t)}$ in terms of the fuzzy
spherical harmonics defined in (\ref{fuzzy spherical harmonics}):
\begin{align}
\phi^{(s,t)}=\sum_{J=|j_s-j_t|}^{j_s+j_t}\sum_{m=-J}^J\phi^{(s,t)}_{Jm}
\otimes \hat{Y}_{Jm(j_sj_t)},
\label{mode expansion in terms of fuzzy spherical harmonics}
\end{align}
where $\phi^{(s,t)}_{Jm}$ is a $k\times k$ matrix. The properties of the fuzzy spherical harmonics are
summarized in appendix B.
We will see shortly that $j_s-j_t=(s-t)/2$ is identified with $\tilde{m}$ as in the case
of the $\phi^3$ theory on $S^1$.
Using the modes, (\ref{reduced model of phi^3 theory}) is expressed as
\begin{align}
S_r=&\frac{vn}{g^2}\left[ \frac{\mu^2}{2}\sum_{s,t}
\sum_{Jm}(-1)^{m-(j_s-j_t)}(J(J+1)+\xi^2)\mbox{tr}(\phi^{(s,t)}_{Jm}\phi^{(t,s)}_{J-m}) \right.
\nonumber\\
&\qquad\left.+\frac{\mu}{3}\sum_{s,t,u}\sum_{J_1m_1J_2m_2J_3m_3}
\hat{{\cal C}}_{J_1m_1(j_sj_t)\:J_2m_2(j_tj_u)\:J_3m_3(j_uj_s)}
\mbox{tr}(\phi^{(s,t)}_{J_1m_1}\phi^{(t,u)}_{J_2m_2}\phi^{(u,s)}_{J_3m_3})\right],
\end{align}
where $\hat{{\cal C}}_{J_1m_1(j_sj_t)\:J_2m_2(j_tj_u)\:J_3m_3(j_uj_s)}$ is defined in (\ref{Chat}),
and tr stands for the trace over $k\times k$ matrices.

Applying the rule (\ref{rule}),
we obtain the observable in the reduced model corresponding to (\ref{correlation function on S^3}):
\begin{align}
\left\langle \frac{1}{N}\mbox{Tr}(G_1^{-1}\phi G_1G_2^{-1}\phi G_2)\right\rangle_r, 
\label{correlation function in reduced model}
\end{align}
where $\langle\cdots\rangle_r$ stands for the VEV in the reduced model.
By using (\ref{mode expansion in terms of fuzzy spherical harmonics}), 
(\ref{action of SU(2) on fuzzy spherical harmonics}) 
and (\ref{orthonormality condition for fuzzy spherical harmonics}), we calculate 
(\ref{correlation function in reduced model}):
\begin{align}
&\frac{1}{\nu}\sum_{s,t}\sum_{JmJ'm'}
\left\langle\frac{1}{k}\mbox{tr}(\phi^{(s,t)}_{Jm}\phi^{(t,s)}_{J'm'})\right\rangle_r 
R_J(g_2g_1^{-1})_{m''m}\frac{1}{n}\mbox{tr}(\hat{Y}_{Jm''(j_sj_t)}\hat{Y}_{J'm'(j_tj_s)})\nonumber\\
&=\frac{1}{\nu}\sum_{s,t}\sum_{Jmm'}(-1)^{m'-(j_t-j_s)}R_J(g_2g_1^{-1})_{-m'm}
\left\langle\frac{1}{k}\mbox{tr}(\phi^{(s,t)}_{Jm}\phi^{(t,s)}_{Jm'})\right\rangle_r.
\label{calculation of correlation function in reduced model}
\end{align}
We calculate again the diagram in Fig.\ref{planar} which appears in the perturbative expansion 
of $\langle\frac{1}{k}\mbox{tr}(\phi^{(s,t)}_{Jm}\phi^{(t,s)}_{Jm'})\rangle_r$:
\begin{align}
\sum_{u}\sum_{Jmm'J_2m_2J_3m_3}
&\frac{(-1)^{m+m'}}{\mu^6}\left(\frac{g^2N}{n^2\nu v}\right)^2
\left(\frac{1}{J(J+1)+\xi^2}\right)^2
\frac{(-1)^{m_2-(j_s-j_u)}}{J_2(J_2+1)+\xi^2}\frac{(-1)^{m_3-(j_u-j_t)}}{J_3(J_3+1)+\xi^2} \nonumber\\
&\times\hat{{\cal C}}_{J-m(j_tj_s)\:J_2m_2(j_sj_u)\:J_3m_3(j_uj_t)}
\hat{{\cal C}}_{J-m'(j_sj_t)\:J_3-m_3(J_tj_u)\:J_2m_2(j_uj_s)}.
\label{diagram in reduced model}
\end{align}

Comparison of (\ref{diagram in phi^3 theory}) with (\ref{diagram in reduced model})
leads us to put
\begin{align}
v=\frac{V_{S^3}}{n^2\nu},
\label{v}
\end{align}
and take a limit 
\begin{align}
k\rightarrow\infty, \;\; \nu\rightarrow\infty,\;\; n/\nu\rightarrow\infty
\;\; \mbox{with} \;\; g^2N \;\; \mbox{fixed}.
\label{limit}
\end{align}
Note again that (\ref{limit}) implies (\ref{planar limit}).
Then, in the limit (\ref{limit}), we can make an identification
$\tilde{m}=j_s-j_t,\;\tilde{m}_2=j_s-j_u,\; \tilde{m}_3=j_u-j_t$ as anticipated
and show by using (\ref{6-j symbol and Clebsch-Gordan}) that
\begin{align}
\hat{{\cal C}}_{J-m(j_tj_s)\:J_2m_2(j_sj_u)\:J_3m_3(j_uj_t)} &\rightarrow 
{\cal C}_{J-m-\tilde{m}\:J_2m_2\tilde{m}_2\:J_3m_3\tilde{m}_3} \nonumber\\
\hat{{\cal C}}_{J-m'(j_sj_t)\:J_3-m_3(J_tj_u)\:J_2m_2(j_uj_s)} &\rightarrow
{\cal C}_{J-m'\tilde{m}\:J_3-m_3-\tilde{m}_3\:J_2m_2-\tilde{m}_2} \ .
\label{limit of 6-j symbol}
\end{align}
We therefore find that (\ref{diagram in reduced model}) agrees with (\ref{diagram in phi^3 theory}).
As before, this correspondence holds for
all the planar diagrams such as Fig.\ref{planar}, while
it does not hold for the non-planar diagrams such as Fig.\ref{nonplanar}.
However, because of $k\rightarrow\infty$, only the planar diagrams
contribute to (\ref{calculation of correlation function in reduced model}).
Note also that the UV/IR mixing on fuzzy spheres does not exist in the planar contribution, so
non-commutativity does not remain in the $n\rightarrow\infty$ limit.
Furthermore, while the summation over $s$ and $t$ 
in (\ref{calculation of correlation function in reduced model}) is redundant compared to
the summation over $\tilde{m}$ in (\ref{calculation of correlation function in phi^3 theory}),
this redundancy is canceled by
a factor $1/\nu$ in (\ref{calculation of correlation function in reduced model}).
Thus we conclude that (\ref{correlation function in reduced model}) in the limit (\ref{limit})
agrees with (\ref{correlation function on S^3}) in the planar limit.

As in the case of the $\phi^3$ theory on $S^1$, 
one can easily show that counterparts of 
(\ref{correspondence of general correlation functions}) and (\ref{correspondence of free energies on S^1})
hold in the limit (\ref{limit}):
\begin{align}
\left\langle \frac{1}{N}\mbox{Tr}(G_1^{-1}\phi G_1G_2^{-1}\phi G_2\cdots G_r^{-1}\phi G_r)\right\rangle_r
=\left\langle \frac{1}{N}\mbox{Tr}(\phi(g_1)\phi(g_2)\cdots\phi(g_r))\right\rangle.
\label{correspondence of correlation functions}
\end{align}
and
\begin{align}
\frac{F_r}{N^2v}=\frac{F}{N^2V_{S^3}},
\label{correspondence of free energies}
\end{align}
Thus, in the limit (\ref{limit}), the reduced model (\ref{reduced model of phi^3 theory}) with (\ref{v}) retrieves 
the planar limit of the original theory (\ref{phi^3 theory})\footnote{Another large-$N$ reduction on $S^3$
was developed in \cite{Kawai:2009vb}, where the planar limit of the original theory is retrieved by setting
$n=(\nu+1)/2$ and $k_s=(2j_s+1)k$ in
(\ref{reducible representation}) instead of (\ref{parameters}) 
and taking the limit in which $\nu\rightarrow\infty$ and $k\rightarrow\infty$
with $v=V_{S_3}k/N$. }.

What we have done is understood as follows. 
$S^3$ is viewed as an $S^1$-bundle over $S^2$.
The momenta along the $S^1$ is given by $\tilde{m}$ in (\ref{harmonic expansion}).
By making the Kaluza-Klein (KK) reduction along the $S^1$, one obtains the KK modes on $S^2$
with the KK momenta given by $\tilde{m}$.
The KK mode with the KK momentum $\tilde{m}$ behaves as in a situation that
the monopole with the monopole charge $\tilde{m}$ is located at the origin of the $S^2$.
Reflecting the angular momentum possessed by the monopole,
the angular momentum $J$ of the KK mode on $S^2$ is restricted to $|\tilde{m}|\leq J$.
The identification $\tilde{m}=j_s-j_t=(s-t)/2$ 
matches the range of $J$ in (\ref{mode expansion in terms of fuzzy spherical harmonics})
in the $n\rightarrow\infty$ limit.
Actually, the monopole spherical harmonics for the monopole charge $q$ is regularized by
the fuzzy spherical harmonics which is an $(l+q)\times l$ 
rectangular matrix \cite{Ishii:2008ib,Ishiki:2006yr,Ishii:2008tm}.
The relation (\ref{limit of 6-j symbol}) reflects this fact.
We obtain the base space $S^2$ through the continuum limit of the fuzzy sphere
and the $S^1$ fiber through the mechanism of the large-$N$ reduction on $S^1$ explained in the previous subsection.
$n$ plays the role of the UV cutoff on $S^2$, while $\nu$ plays the role of the UV cutoff on $S^1$.

\subsection{${\cal N}=4$ SYM on $R\times S^3$ from PWMM}
We first see that ${\cal N}=4$ SYM on $R^4$ at a conformal point, where the 
VEVs
of all the scalar fields vanish, is equivalent to ${\cal N}=4$ SYM on $R\times S^3$ through 
the conformal map.
In this subsection, for simplicity, we suppress the terms including fermion fields.
Then, the action of ${\cal N}=4$ $U(N)$ SYM on $R^4$ takes the form
\begin{align}
S_{YM}=\frac{1}{g^2}\int d^4x \mbox{Tr}
\left(F_{\mu\nu}^2+\frac{1}{2}(D_{\mu}\phi_m)^2
-\frac{1}{4}[\phi_m,\phi_n]^2\right),
\label{action of N=4 SYM on R^4}
\end{align}
where $D_{\mu}=\partial_{\mu}+i[A_{\mu}, \;\;]$. There are six scalars $\phi_m$, 
and for later convenience we make $m$ and $n$ run from 4 to 9.
This theory possesses the $PSU(2,2|4)$ superconformal symmetry with 32 supercharges.
Applying the Weyl transformation defined by
\begin{align}
& A_{\mu} \rightarrow A_{\mu} \ , \nonumber\\
& \phi_m \rightarrow e^{-\frac{\rho(x)}{2}}\phi_m \ , \nonumber\\
&\delta_{\mu\nu}\rightarrow g_{\mu\nu}=e^{\rho(x)}\delta_{\mu\nu} \ .
\label{Weyl transformation}
\end{align}
to (\ref{action of N=4 SYM on R^4}) yields ${\cal N}=4$ SYM on a curved space endowed with a metric $g_{\mu\nu}$:
\begin{align}
S_{YM}=\frac{1}{g^2}\int d^4x \sqrt{g} \mbox{Tr}
\left(\frac{1}{4}g^{\mu\lambda}g^{\nu\rho}F_{\mu\nu}F_{\lambda\rho}+\frac{1}{2}g^{\mu\nu}D_{\mu}\phi_mD_{\nu}\phi_m
+\frac{1}{12}R\phi_m^2-\frac{1}{4}[\phi_m,\phi_n]^2\right) \ ,
\label{action of N=4 SYM on curved space}
\end{align}
where $R$ is the Ricci scalar constructed from $g_{\mu\nu}$.
$R^4$ is transformed to $R\times S^3$ by the Weyl transformation.
In fact, if one starts with the metric of $R^4$ in 
the polar coordinate and rewrites it as 
\begin{align}
ds_{R^4}^2=dr^2 +r^2 d\Omega_3^2
          =e^{\mu \tau}\left(d\tau^2
            +\left(\frac{2}{\mu}\right)^2 d\Omega_3^2 \right),
\label{metric}
\end{align}
one obtains the metric of $R \times S^3$ 
up to the conformal factor $e^{\mu \tau}$. Namely, $\rho=-\mu \tau$ 
in this case.
Here we have changed the coordinate from $r$ to 
$\tau \equiv \frac{2}{\mu} \log \left( \frac{\mu r}{2} \right)$ 
where $2/\mu$ is the radius of the resultant $S^3$ as we have adopted so far.
Thus ${\cal N}=4$ SYM on $R^4$ at the conformal point is equivalent to 
${\cal N}=4$ SYM on $R\times S^3$.

Next, using the equations in appendix A, we rewrite the action (\ref{action of N=4 SYM on curved space}) 
for $R\times S^3$ 
in terms of the Killing vector. We expand the gauge field on $S^3$ as
\begin{align}
A=X_ie^i,
\end{align}
where $e^i\; (i=1,2,3)$ are defined in (\ref{right invariant 1-form}),  
and put $\phi_m=X_m$.
By using (\ref{Maurer-Cartan}), we rewrite (\ref{action of N=4 SYM on curved space}) in terms of $X_i$ and $X_m$:
\begin{align}
S_{YM}&=\frac{V_{S^3}}{g^2}\int d\tau\frac{d\Omega_3}{2\pi^2}
\mbox{Tr} \left( \frac{1}{2}(D_{\tau}X_i-i\mu{\cal L}_iA_{\tau})^2
+\frac{1}{2}\left(\mu X_i+i\epsilon_{ijk}(\mu{\cal L}_jX_k+\frac{1}{2}[X_j,X_k])\right)^2 \right. \nonumber\\
&\qquad\qquad\qquad\qquad\;\;\;\left.+\frac{1}{2}(D_{\tau}X_m)^2-\frac{1}{2}(\mu{\cal L}_iX_m+[X_i,X_m])^2
+\frac{\mu^2}{8}X_m^2-\frac{1}{4}[X_m,X_n]^2\right),
\end{align}
where $D_{\tau}=\partial_{\tau}+i[A_{\tau},\;]$.
Applying the rule (\ref{rule}) to the above action, we obtain a reduced model of 
${\cal N}=4$ SYM on $R\times S^3$:
\begin{align}
S_{YM,r}&=\frac{v}{g^2}\int d\tau
\mbox{Tr} \left( \frac{1}{2}(D_{\tau}X_i-i\mu [L_i,A_{\tau}])^2
+\frac{1}{2}\left(\mu X_i+i\epsilon_{ijk}(\mu[L_j,X_k]+\frac{1}{2}[X_j,X_k])\right)^2 \right. \nonumber\\
&\qquad\qquad\qquad\;\;\left.+\frac{1}{2}(D_{\tau}X_m)^2-\frac{1}{2}(\mu[L_i,X_m]+[X_i,X_m])^2
+\frac{\mu^2}{8}X_m^2-\frac{1}{4}[X_m,X_n]^2\right).
\label{reduced model of N=4 SYM on RxS^3}
\end{align}
Note that the matrices still depend on $\tau$.

Here we can absorb $\mu L_i$ into $X_i$:
\begin{align}
\mu L_i+X_i  \rightarrow X_i.
\label{absorption}
\end{align}
In other words, we can regard $\mu L_i$ as a background of $X_i$.
Note that one can not perform such an absorption in the case of the 
$\phi^3$ theory in the previous subsection.
The resulting action is nothing but the plane wave matrix model \cite{Berenstein:2002jq}:
\begin{align}
S_{\rm PW}
= \frac{1}{g_{PW}^2}
\int
d\tau \, \mbox{Tr} &
\left(\frac{1}{2}(D_{\tau}X_M)^2-\frac{1}{4}[X_M,X_N]^2
+\frac{\mu^2}{2}X_i^2
+\frac{\mu^2}{8}X_m^2 +i\mu\epsilon_{ijk}X_iX_jX_k
\right) \ ,
\label{pp-action}
\end{align}
where $M$ runs from 1 to 9 and we put
\begin{align}
\frac{g^2}{v}=g_{PW}^2.
\end{align}
This action is obtained by dimensionally reducing ${\cal N}=4$ SYM on $R\times S^3$ over $S^3$ \cite{Kim:2003rza,Lin}.

The PWMM possesses the $SU(2|4)$ supersymmetry, which includes
16 supercharges.
The PWMM possesses many discrete vacua given by $X_i=\mu L_i$, where 
$L_i$ is given in (\ref{reducible representation}), and they are degenerate.
There are no classical moduli, and these vacua are classically stable.
Furthermore, all of these vacua preserve the $SU(2|4)$ supersymmetry, and
the theories around these vacua are massive.
Hence they are also quantum mechanically stable at least at perturbative level \cite{Dasgupta:2002hx}.
We pick up a particular one specified by (\ref{parameters}) from these vacua 
and take the limit (\ref{limit}) to obtain the reduced model of ${\cal N}=4$ SYM on $R\times S^3$.
In this limit, this vacuum is stable even at non-perturbative level, because tunneling to other vacua by an instanton effect
is suppressed by the $k\rightarrow\infty$ limit.
The global gauge invariance (depending only on $\tau$) 
of the PWMM is translated to
the local gauge invariance of ${\cal N}=4$ SYM.
The $SU(2|4)$ symmetry is expected to enhance to the $PSU(2,2|4)$ symmetry in the continuum limit.
Thus the reduced model (\ref{reduced model of N=4 SYM on RxS^3}) retrieves the planar limit of
${\cal N}=4$ SYM on $R\times S^3$ 
as the reduced model (\ref{reduced model of phi^3 theory}) does
the planar limit of (\ref{phi^3 theory}).

Note that obtaining a reduced model of ${\cal N}=4$ SYM on $R^4$ is quite non-trivial.
For instance, dimensionally reducing it to $R$ yields
the D0-brane effective theory
or the Matrix theory \cite{Banks:1996vh}, which is obtained by putting $\mu=0$ in (\ref{pp-action}).
This theory possesses infinitely many continuous vacua, which are characterized by
$[X_M,X_N]=0$. All of these vacua preserve sixteen supercharges as well.
However, there are one-dimensional massless fields around these vacua
at least perturbatively, so whether each vacuum is stable at quantum level heavily depends on dynamics.
Thus one may not naively expect 
that a theory around a vacuum where
$X_i$ are simultaneously diagonal and their eigenvalues distribute uniformly in $R^3$
retrieves the planar limit of ${\cal N}=4$ SYM on $R^4$.

We summarize the prescription of the large-$N$ reduction for ${\cal N}=4$ SYM on $R\times S^3$.
We expand the PWMM (\ref{pp-action}) around the vacuum $X_i=\mu L_i$ with (\ref{reducible representation})
and (\ref{parameters}), and take the limit
\begin{align}
k\rightarrow\infty,\;\;\nu\rightarrow\infty,\;\;n/\nu\rightarrow\infty
\;\;\mbox{with}\;\; \frac{g_{PW}^2k}{n}=\frac{g^2N}{V_{S^3}} \;\;\mbox{fixed}.
\label{continuum limit for N=4}
\end{align}
Then, the planar limit of ${\cal N}=4$ SYM on $R\times S^3$ with the 't Hooft coupling $g^2N$ is retrieved.
The correspondence is explicitly given in (\ref{correspondence of correlation functions})
and (\ref{correspondence of free energies}).
Thus the planar limit of ${\cal N}=4$ SYM on $R\times S^3$ is regularized by the PWMM. 
This regularization preserves the $SU(2|4)$ symmetry including sixteen supercharges and the gauge symmetry.
This number of the preserved supercharges is optimal.
In the next section, we will see the correspondence between Wilson loops.

\section{Wilson loops}
\setcounter{equation}{0}
In this section, we examine the correspondence between Wilson loops in ${\cal N}=4$ SYM and 
its reduced model we reviewed in the previous section.
In section 3.1, we review the Wilson loops considered in the gauge theory side
in the context of the AdS/CFT correspondence.
In section 3.2, we construct the corresponding Wilson loop in the reduced model.

\subsection{Wilson loop in the AdS/CFT correspondence}
The Wilson loop considered in the AdS/CFT correspondence
takes the form
\begin{align}
W(C)=\frac{1}{N}{\rm Tr}{\cal P} \exp{ \left(i\int_0^1 ds \{\dot{x}^{\mu}(s)
A_{\mu}(x(s)) 
+i|\dot{x}(s)| \Theta^m(s) \phi_m(x(s)) \} \right)},
\label{susy Wilson loop}
\end{align}
where the function $x^{\mu}(s): [0,1] \rightarrow C$ 
specifies the contour $C$, and $\Theta^m(s)$ 
satisfies $\delta_{mn}\Theta^m\Theta^n=1$.
The Wilson loop is invariant under the Weyl transformation (\ref{Weyl transformation}), 
namely takes
the same form both in $R^4$ and $R\times S^3$.
The contour $C$ on $R^4$ is mapped to the corresponding contour $C'$ on $R\times S^3$
given by the coordinate transformation used in (\ref{metric}), and
the following equality holds:
\begin{align}
\langle W(C) \rangle_{R^4} = \langle W(C') \rangle_{R\times S^3}.
\label{map for Wilson loop}
\end{align}

We next consider the supersymmetric property of the 
Wilson loop on $R\times S^3$ for later convenience.
Its infinitesimal variation under the supersymmetry transformation
is proportional to
$(\dot{x}^{\mu}\Gamma_{\mu}+i|\dot{x}|\Theta^m\Gamma_m)\epsilon(x)$
where $\mu =\tau, \theta, \varphi, \psi$ and 
$\epsilon(x)$ is a ten-dimensional Killing spinor on 
$R\times S^3$ satisfying the Killing spinor equation 
\cite{Okuyama:2002zn,Ishiki:2006rt},
\begin{align}
\nabla_a \epsilon(x)=\pm
\frac{\mu}{4}\Gamma_a \Gamma^1\Gamma^2\Gamma^3\epsilon(x).
\label{Killing spinor equation}
\end{align}
Here $a=\tau,1,2,3$ is the local Lorentz index. 
This equation is solved by
\begin{align}
\epsilon_1(x)=e^{\frac{\mu}{4}\Gamma^{\tau 123}\tau}\eta_1, 
\;\;
\epsilon_2(x)=e^{-\frac{\mu}{4}\Gamma^{\tau 123}\tau}
e^{-\frac{1}{2}\Gamma^{12}\varphi}
e^{-\frac{1}{2}\Gamma^{31}\theta}
e^{-\frac{1}{2}\Gamma^{12}\psi}\eta_2,
\end{align}
where $\epsilon_1$ and $\epsilon_2$ are 
the solutions to  
the upper and the lower signs of (\ref{Killing spinor equation}),
respectively, and $\eta_1$ and $\eta_2$ are constant spinors.
If the Wilson loop operator is invariant under
some of the supersymmetries, 
there exist some non-zero components of $\eta_{1,2}$ 
such that the following two equations hold,
\begin{align}
&(\dot{x}^{\mu}\Gamma_{\mu}+i|\dot{x}|\Theta^m\Gamma_m)
e^{ \frac{\mu}{4}\Gamma^{\tau 123}\tau}\eta_1 =0 ,\nonumber\\
&(\dot{x}^{\mu}\Gamma_{\mu}+i|\dot{x}|\Theta^m\Gamma_m)
e^{-\frac{\mu }{4}\Gamma^{\tau 123}\tau}
e^{-\frac{1}{2}\Gamma^{12}\varphi}
e^{-\frac{1}{2}\Gamma^{31}\theta}
e^{-\frac{1}{2}\Gamma^{12}\psi}\eta_2=0.
\label{BPS equation}
\end{align}
Introducing local projection operators as
\begin{align}
P_{1\pm}&=\frac{1}{2}e^{-\frac{\mu }{4}\Gamma^{\tau 123}\tau}
(1\pm\frac{i\dot{x}^{\mu}}{|\dot{x}|}\Theta^m\Gamma_{\mu m})
e^{ \frac{\mu }{4}\Gamma^{\tau 123}\tau},
\nonumber\\
P_{2\pm}&=\frac{1}{2}
e^{\frac{1}{2}\Gamma^{12}\psi}
e^{\frac{1}{2}\Gamma^{31}\theta}
e^{\frac{1}{2}\Gamma^{12}\varphi}
e^{\frac{\mu}{4}\Gamma^{\tau 123}\tau}
(1\pm\frac{i\dot{x}^{\mu}}{|\dot{x}|}\Theta^m\Gamma_{\mu m})
e^{-\frac{\mu}{4}\Gamma^{\tau 123}\tau}
e^{-\frac{1}{2}\Gamma^{12}\varphi}
e^{-\frac{1}{2}\Gamma^{31}\theta}
e^{-\frac{1}{2}\Gamma^{12}\psi},
\end{align}
one can rewrite (\ref{BPS equation}) simply as
\begin{align}
P_{1+}\eta_1=0,\;\;P_{2+}\eta_2=0.
\label{BPS equation 2}
\end{align}
The number of independent non-zero components 
of $\eta_1$ and $\eta_2$ which satisfy (\ref{BPS equation 2})
(or equivalently (\ref{BPS equation}))
is just the number of supersymmetries preserved by the
insertion of the operator.

As an example, 
let us consider the case where $\dot{x}^{\mu}(s)$ and
$\Theta(s)$ are
constant in $s$ and the contour does not extend to the 
$\tau$ direction.
In this case, the projection operator $P_1$ is  
a constant operator on the contour, so that 
the half components of $\eta_1$ which are
projected by $P_{1+}$ into 0 can be non-zero constants.
Therefore, the operator preserves 
at least 8 supersymmetries (1/4 BPS).
We can also consider a special case of this example
in which the path is given by a great circle 
on $S^3$ at a fixed value of $\tau$,
\begin{align}
C:\;(\tau(s),\theta(s), \varphi(s), \psi(s))=(\tau_0,0,0,4\pi s),
\label{path}
\end{align}
where $\tau_0$ is a constant.
In addition to the above mentioned 8 supersymmetries, 
another 8 supercharges are preserved in this case
because $P_2$ is also a constant operator on the contour (\ref{path})
as well as $P_1$.
Hence, the Wilson loop on the great circle (\ref{path}) 
is a half-BPS operator.

The contour (\ref{path}) is mapped to the circular loop, whose center is located at 
the origin on $R^4$, by the Weyl transformation (\ref{Weyl transformation}).
For the circular contour with the unit radius,
$x^{\mu}(s)$ can be parametrized as 
\begin{align}
\{x^{\mu}(s)\}=(\cos (2\pi s), \sin (2\pi s), 0,0).
\label{circle on R^4}
\end{align}
Because of (\ref{map for Wilson loop}), the VEV of 
the Wilson loop with the contour (\ref{path}) on $R\times S^3$
coincides with the VEV of the Wilson loop with the contour 
(\ref{circle on R^4}) on $R^4$. 

The VEV of the BPS circular Wilson loop
on $R^4$ can be computed 
including all orders in the perturbative expansion
in the planar limit
\cite{Erickson:2000af,Drukker:2000rr}.
Let us review the result of this computation.
The VEV of (\ref{susy Wilson loop}) for the path 
(\ref{circle on R^4}) turns out to be given by the
following VEV in the Gaussian matrix model,
\begin{align}
\langle W({\rm circle})\rangle 
= \Big\langle \frac{1}{k} {\rm Tr} \exp(M) \Big\rangle
\equiv \frac{1}{Z}\int {\cal D}M \frac{1}{k} {\rm Tr} \exp(M)
\exp \left(-\frac{k}{\lambda} {\rm Tr} M^2 \right)
= \sqrt{\frac{2}{\lambda}} I_1(\sqrt{2\lambda}),
\label{ESZ result}
\end{align}
where $\lambda$ is the 't Hooft coupling\footnote{Our convention of 
the 't Hooft coupling is slightly different from that in 
\cite{Erickson:2000af}.
The relation is given by
$\lambda_{{\rm ours}}=\lambda_{ESZ}/2$.} and $I_{1}$ is a Bessel function.
The strong coupling expansion of this result
reproduces the correct coupling behavior $e^{\sqrt{\lambda}}$ 
observed in the gravity side.

\subsection{Wilson loop in the reduced model}
By using the relation (\ref{correspondence of correlation functions}),
we can construct an operator in the reduced model 
(\ref{reduced model of N=4 SYM on RxS^3})
corresponding to the Wilson loop (\ref{susy Wilson loop}):
\begin{align}
\hat{W}(C)=\frac{1}{N}
{\rm Tr}&\left[ {\cal P}
\exp \left\{ i\int^1_0 ds
\left( 
\dot{x}^{\tau}(s)G^{-1}(s)A_{\tau}(\tau(s))G(s) \right.\right.\right. \nonumber\\
&+\dot{x}^{\bar{\mu}}(s)e^i_{\bar{\mu}}(x^{\bar{\mu}}(s))
G^{-1}(s)X_i(\tau(s))G(s) \nonumber\\
&\left.\left.\left.+i|\dot{x}(s)|\Theta^m(s)G^{-1}(s)X_m(\tau(s)) G(s)
\right)
\right\}
\right], 
\label{Wilson loop in PWMM}
\end{align}
where $\bar{\mu}=\theta,\varphi,\psi$ and
$e^i$ is the right invariant 1-form on $S^3$ defined in 
(\ref{right invariant 1-form}).
$G(s)$ is given by
\begin{align}
G(s)=P\exp \left[i\int_0^s ds' \dot{x}^{\bar{\mu}}(s')
e^i_{\bar{\mu}}(x^{\bar{\mu}}(s'))\mu L_i \right],
\end{align}
where $L_i$ is given by (\ref{reducible representation}) and (\ref{parameters}).
Note that $G(1)=1$. 
By referring to an argument in \cite{Ishibashi:1999hs}, we can show
that (\ref{Wilson loop in PWMM}) is equal to \cite{Ishii:2007sy}
\begin{align}
\hat{W}(C)=\frac{1}{N}
{\rm Tr}&\left[ {\cal P}
\exp \left\{ i\int^1_0 ds
\left( 
\dot{x}^{\tau}(s)A_{\tau}(\tau(s)) 
+\dot{x}^{\bar{\mu}}(s)e^i_{\bar{\mu}}(x^{\bar{\mu}}(s))
(\mu L_i+X_i(\tau(s))) \right.\right.\right.\nonumber\\
&\left.\left.\left.+i|\dot{x}(s)|\Theta^m(s) X_m(\tau(s))
\right)
\right\}
\right].
\label{Wilson loop in PWMM 2} 
\end{align}
In fact, (\ref{Wilson loop in PWMM}) can be viewed as $1/N$ times the trace of the time evolution
kernel for a time-dependent Hamiltonian
\begin{align}
H(s)=&-\left\{ \dot{x}^{\bar{\mu}}(s)e^i_{\bar{\mu}}(x^{\bar{\mu}}(s))
\mu L_i+\dot{x}^{\tau}(s)A_{\tau}(\tau(s)) 
+\dot{x}^{\bar{\mu}}(s)e^i_{\bar{\mu}}(x^{\bar{\mu}}(s))X_i(\tau(s)) 
\right.
\nonumber\\
& \left. 
+i|\dot{x}(s)|\Theta^m(s) X_m(\tau(s))
\right\}.
\end{align}
If one regards the first term as a free part and the others as an 
interaction part
and switches the picture to the interaction one, one obtains (\ref{Wilson loop in PWMM}).
Note that applying (\ref{absorption}) to (\ref{Wilson loop in PWMM 2}) results in
the dimensional reduction of (\ref{susy Wilson loop}) from $R\times S^3$ to $R$.
In other words, we expand $X_i$ around $\mu L_i$ in the dimensionally reduced Wilson loop.
Such a correspondence for the Wilson loops is seen commonly in the large-$N$ reduction.

$\mbox{}$From (\ref{correspondence of correlation functions}), we expect 
\begin{align}
\langle \hat{W}(C) \rangle_r=\langle W(C) \rangle.
\end{align}
In the next section, we verify this relation 
in the case of the half-BPS Wilson loop with the contour $C$ given in (\ref{path}).

\section{VEV of the half-BPS Wilson loop from PWMM}
\setcounter{equation}{0}
In this section, we calculate VEV of the operator 
(\ref{Wilson loop in PWMM 2}) with the contour (\ref{path}).

For the calculation of the circular BPS Wilson loop in SYM on $R^4$, 
it is convenient to take the Feynman gauge where the computation is simplified 
such that the non-ladder diagrams cancel out \cite{Erickson:2000af}.
One can expect that this simplification also occurs in the PWMM if 
one takes a particular gauge which corresponds to the Feynman gauge in the 
continuum limit. This leads us to take this particular gauge 
in the PWMM for the following computation. 

The corresponding gauge fixing term on $R\times S^3$ can be obtained 
by applying the Weyl transformation.
Although the original action is invariant under the Weyl 
transformation (\ref{Weyl transformation}), the gauge fixing term is not 
invariant in general. 
In fact, if one takes the Feynman gauge on $R^4$, 
the gauge fixing term is transformed as
\begin{align}
\frac{1}{2}\sqrt{g^{R^4}}\left( g_{R^4}^{\mu\nu} 
\nabla_{\mu}^{R^4}A_{\nu}^{R^4}\right)^2
&=
\frac{1}{2\sqrt{g^{R^4}}}\left\{ 
\partial_{\mu}(\sqrt{g^{R^4}}g_{R^4}^{\mu\nu}A_{\nu}^{R^4})
\right\}^2 \nonumber\\
&=
\frac{e^{-2\mu\tau}}{2\sqrt{g^{R\times S^3}}}\left\{ 
\partial_{\mu}(e^{\mu\tau}
\sqrt{g^{R\times S^3}}g_{R\times S^3}^{\mu\nu}
A_{\nu}^{R\times S^3}) \right\}^2 \nonumber\\
&=
\frac{1}{2}\sqrt{g^{R\times S^3}}\left( g_{R\times S^3}^{\mu\nu} 
\nabla_{\mu}^{R\times S^3}A_{\nu}^{R\times S^3}
+\mu A_{\tau}^{R\times S^3} \right)^2.
\label{gauge fixing}
\end{align}
Hence, the corresponding gauge fixing term on $R\times S^3$ should be
given by the last line of (\ref{gauge fixing}).
We then apply the reduction rule (\ref{rule}) to the above 
gauge fixing term plus appropriate ghost terms to obtain those in 
the PWMM,
\begin{align}
S_{gf+gh}&=\frac{1}{g_{PW}^2}
\int d\tau {\rm Tr} \Big\{ 
\frac{1}{2}(\partial_{\tau}A_{\tau}+i\mu [L_i,X_i]+\mu A_{\tau})^2
\nonumber\\
& \;\;\; -i\bar{c}(\partial_{\tau}D_{\tau}+\mu D_{\tau})c
+\mu \bar{c}[L_i,i\mu [L_i,c]+i[c,X_i]]
\Big\},
\label{gauge fixing plus ghost}
\end{align}
where $c$ and $\bar{c}$ are the ghost fields.

We add (\ref{gauge fixing plus ghost}) to the original action of 
the PWMM
expanded around the background (\ref{reducible representation})
with the parameters given by (\ref{parameters}) and 
construct the Feynman rule of this theory.
We make a mode expansion of the matrices as in
(\ref{mode_expansion_in_PWMM}) and
read off the propagators for each momentum mode as

\vspace{-0.5cm}
{ \small
\begin{align}
&\langle x_{Jm\rho}^{(s,t)}(p)_{ij} x_{J'm'\rho'}^{(s',t')}(p')_{kl}\rangle
\n
&=
\begin{cases}
\frac{g^2_{PW}}{n}(-1)^{m-(j_s-j_t)+1}\delta_{JJ'}\delta_{m\,-m'}\delta_{\rho\rho'}
\delta_{st'}\delta_{ts'}\delta_{il}\delta_{jk}
2\pi\delta(p+p')\frac{1}{p^2+{\omega_J^x}^2}
\;\;(\rho\neq0)\\
\frac{g^2_{PW}}{n}
(-1)^{m-(j_s-j_t)+1}\delta_{JJ'}\delta_{m\,-m'}
\delta_{st'}\delta_{ts'}\delta_{il}\delta_{jk}
2\pi\delta(p+p')\frac{p^2+\mu^2(J(J+1)+1)}
{(p^2+\mu^2J^2)(p^2+\mu^2(J+1)^2)}
\;\;(\rho=\rho'=0 )
\end{cases},
\displaybreak[0]\n
&\langle B_{Jm}^{(s,t)}(p)_{ij} B_{J'm'}^{(s',t')}(p')_{kl}\rangle
=
\frac{g^2_{PW}}{n}
(-1)^{m-(j_s-j_t)}\delta_{JJ'}\delta_{m\,-m'}
\delta_{st'}\delta_{ts'}\delta_{il}\delta_{jk}
2\pi\delta(p+p')
{\scriptstyle
\frac{p^2+\mu^2J(J+1)}
{(p^2+\mu^2J^2)(p^2+\mu^2(J+1)^2)}
},
\displaybreak[0]\n
&\langle x_{Jm0}^{(s,t)}(p)_{ij} B_{J'm'}^{(s',t')}(p')_{kl}\rangle
=
\frac{g^2_{PW}}{n}
(-1)^{m-(j_s-j_t)}\delta_{JJ'}\delta_{m\,-m'}
\delta_{st'}\delta_{ts'}\delta_{il}\delta_{jk}
2\pi\delta(p+p')
{\scriptstyle
\frac{i\mu\sqrt{J(J+1)}}
{(p^2+\mu^2J^2)(p^2+\mu^2(J+1)^2)}
},
\displaybreak[0]\n
&\langle c_{Jm}^{(s,t)}(p)_{ij} \bar{c}_{J'm'}^{(s',t')}(p')_{kl}\rangle
=
\frac{g^2_{PW}}{n}
(-1)^{m-(j_s-j_t)}\delta_{JJ'}\delta_{m\,-m'}
\delta_{st'}\delta_{ts'}\delta_{il}\delta_{jk}
2\pi\delta(p+p')
{\scriptstyle
\frac{i}{-p^2+i\mu p -\mu^2J(J+1)}
},
\displaybreak[0]\n
&\langle \phi_{AB,Jm}^{(s,t)}(p)_{ij} \phi_{A'B',J'm'}^{(s',t')}(p')_{kl}
\rangle
=
\frac{g^2_{PW}}{n}
\frac{1}{2}\epsilon_{ABA'B'}
(-1)^{m-(j_s-j_t)}\delta_{JJ'}\delta_{m\,-m'}
\delta_{st'}\delta_{ts'}\delta_{il}\delta_{jk}
2\pi\delta(p+p')
{\scriptstyle
\frac{1}{p^2+{\omega_J^\phi}^2}
},
\displaybreak[0]\n
&\langle \psi_{Jm\kappa}^{A(s,t)}(p)_{ij} \psi_{A',J'm'\kappa'}^{(s',t')\dagger}(p')_{kl}\rangle
=
\frac{g^2_{PW}}{n}
\delta_{JJ'}\delta_{mm'}\delta_{\kappa\kappa'}\delta^A_{A'}
\delta_{ss'}\delta_{tt'}\delta_{il}\delta_{jk}
2\pi\delta(p-p')
{\scriptstyle
\frac{p+\kappa\omega_J^\psi}{p^2+{\omega_J^\psi}^2}
},
\label{propagators in another gauge}
\end{align}
}
\hspace{-2.5mm}
where $\omega_J^x$, $\omega_J^{\phi}$ and 
$\omega_J^{\psi}$ are defined in (\ref{mass}).
We can also read off the interaction vertices 
in this mode expansion.
In particular the vertices used in this section
take the same form as in (\ref{harmonic_expansion_of_S_gauge_int}) 
although a different gauge is taken in 
appendix \ref{Harmonic expansion of PWMM}.

\begin{figure}[tbp]
\begin{center}
\includegraphics[height=5cm, keepaspectratio, clip]{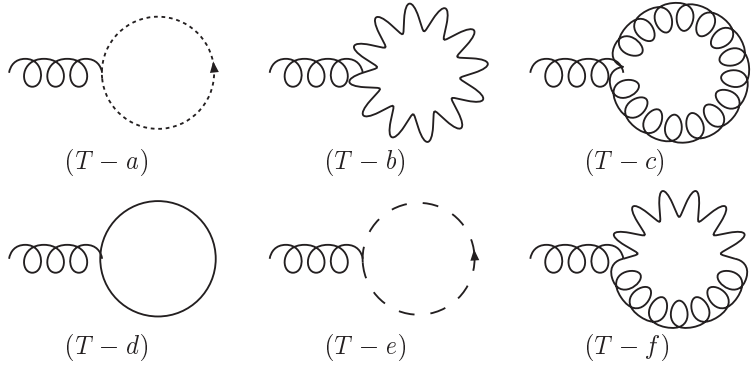}
\end{center}
\caption{Tadpole diagrams. The curly, wavy, 
dotted, solid and dashed lines
represent the propagators of $X_i$, $A_t$, the ghost,
$\Phi_{AB}$ and $\psi^A$ respectively.}
\label{tadpole diagrams}
\end{figure}
In general, the tadpole is expected to vanish as in the
continuum theory.
Let us check it at the 1-loop level in the present gauge.
The only possibly nonzero contribution is
the truncated 1-point function for 
$x_{Jm\rho}^{(s,t)}(p)_{ij}$, where $i,j$ run from 
1 to $k$ and $p$ is
the momentum along the $\tau$ direction.
The 1-point function takes the form,
\begin{align}
2\pi\delta(p)\delta_{st}\delta_{ij}
\delta_{\rho\: -1}\delta_{J0}\delta_{m0}\delta_{ij}\Upsilon^{(s)}.
\label{tpole}
\end{align}
There are six diagrams for the 1-loop correction to 
(\ref{tpole}) as shown in Fig.~\ref{tadpole diagrams}. 
Note that all these diagrams are planar ones.
The diagrams $(T-a)$ and $(T-b)$ completely cancel each other.
The diagram $(T-f)$ vanishes under the integration of $p$
because it is an odd function of $p$.
Below we list the value of $\Upsilon^{(s)}$ for 
each of the remaining diagrams,
\begin{align}
&(T-c) 
= 4
\frac{g^2_{PW}}{\sqrt{n}}
\sum_{t,R}k_t(-1)^{R+j_s+j_t}\sqrt{R(R+1)(2R+1)}
\sixj{1}{R}{R}{j_t}{j_s}{j_s}, \\
&(T-d) 
=12
\frac{g^2_{PW}}{\sqrt{n}}
\sum_{t,R}k_t(-1)^{R+j_s+j_t}\sqrt{R(R+1)(2R+1)}
\sixj{1}{R}{R}{j_t}{j_s}{j_s}, \\
&(T-e) 
=-16
\frac{g^2_{PW}}{\sqrt{n}}
\sum_{t,R}k_t(-1)^{R+j_s+j_t}\sqrt{R(R+1)(2R+1)}
\sixj{1}{R}{R}{j_t}{j_s}{j_s},
\end{align}
where the last factor in the above equations is the $6$-$j$ symbol.
We therefore find that the 1-loop contribution to the tadpole 
in the present gauge is indeed vanishing without taking the 
continuum limit. 

Let us calculate the VEV of the Wilson loop 
(\ref{Wilson loop in PWMM})
with the contour given by a great circle (\ref{path}).
Thanks to the $SO(6)_R$ rotational invariance,
we can set $\Theta^m=\delta^{m4}$ without loss of generality. 
Then, the operator (\ref{Wilson loop in PWMM}) for 
the great circle can be written in a relatively simple form
by substituting (\ref{explicit form of right invariant 1-form})
for the circular contour,
\begin{align}
\hat{W}=\frac{1}{\nu n k}
{\rm Tr}\left(
e^{4\pi i(L_3+H)}
\right),
\end{align}
where we have defined a dimensionless complex matrix $H$ as
\begin{align}
H=\frac{1}{\mu}(X_3+iX_4).
\end{align}
$H$ is expanded in terms of the fuzzy spherical harmonics as
\begin{align}
H^{(s,t)}
&=\frac{1}{\mu}\sum_{R,m}\Bigl(
i C^{R+1m}_{Rm10}
x_{Rm1}^{(s,t)} 
+C^{Rm}_{Rm10}
x_{Rm0}^{(s,t)}
-iC^{R-1m}_{Rm10}
x_{R-1m-1}^{(s,t)} 
 +i(\phi_{14,Rm}^{(s,t)}+\phi_{23,Rm}^{(s,t)})
\Bigr)\otimes \hat{Y}_{Rm}^{(j_sj_t)} \n
&\equiv\sum_{R,m} H^{(s,t)}_{Rm} \otimes \hat{Y}^{(j_sj_t)}_{Rm},
\end{align}
where we have used (\ref{vector and spinor fuzzy spherical harmonics}), (\ref{SU(4) and SO(6)}) and (\ref{mode_expansion_in_PWMM}).
Here the second term ($x_{Rm0}^{(s,t)}$) and the third term 
($x_{R-1m-1}^{(s,t)}$) in the right-hand side of the first line
are summed over $R\geq \frac{1}{2}$ and $R\geq 1$, respectively, 
while the first term ($x_{Rm1}^{(s,t)}$) and the last term 
($\phi_{AB,Rm}^{(s,t)}$) are summed over $R\geq 0$.
Using (\ref{propagators in another gauge}) and the explicit expression of 
the Clebsch-Gordan coefficients
\begin{align}
&C^{R+1m}_{Rm10}=\left[\frac{(R+1+m)(R+1-m)}{(2R+1)(R+1)}\right]^{1/2},
\nonumber\\
&C^{Rm}_{Rm10}=\frac{m}{[R(R+1)]^{1/2}},
\nonumber\\
&C^{R-1m}_{Rm10}=\left[\frac{(R+m)(R-m)}{R(2R+1)}\right]^{1/2},
\end{align}
we can calculate the propagator for $H$. 
In particular, the equal-time propagator is simply given by 
\begin{align}
\langle H_{Rm}^{(s,t)}(\tau)_{ij}H_{R'm'}^{(s',t')}(\tau)_{kl} \rangle
=-\delta_{R0}\delta_{R'0}\delta_{m0}\delta_{m'0}
\delta_{st}\delta_{s't'}\delta_{st'}\delta_{il}\delta_{jk}
\frac{g^2_{PW}}{2\mu^3n}.
\label{propagator for H}
\end{align}
Namely, only the zero mode contributes to the propagator of $H$
in the present gauge.
We expand the VEV of the Wilson loop as
\begin{align}
\langle \hat{W} \rangle 
&=
\frac{1}{\nu nk}\sum_{q=0}^{\infty}
\frac{1}{q!} \left( \frac{d^q}{da^q}
{\rm Tr} \langle e^{4\pi i (L_3 +a H)} 
\rangle \right)_{a=0} \nonumber\\
&=
\frac{1}{\nu n k}\sum_{q=0}^{\infty}
\int_0^1 d\alpha_1\int_0^1 d\alpha_2 \cdots \int_0^1 d\alpha_{q+1}
\delta (1-\alpha_1-\alpha_2-\cdots -\alpha_{q+1}) \nonumber\\
&{} \hspace{2cm} \times (4\pi i)^q
{\rm Tr} \langle e^{4\pi i \alpha_1 L_3}H
e^{4\pi i \alpha_2 L_3}H \cdots e^{4\pi i \alpha_{q} L_3}H
e^{4\pi i \alpha_{q+1} L_3}
\rangle ,
\end{align}
where we have used the generalized Feynman formula 
proven in appendix \ref{Generalization of the Feynman formula}
to show the second equality.
Then we obtain
\begin{align}
\langle \hat{W} \rangle 
&=
\frac{1}{\nu nk}\sum_{q=0}^{\infty}
\sum_{\{s_i\}}
\int_0^1 d\alpha_1\int_0^1 d\alpha_2 \cdots \int_0^1 d\alpha_{q+1}
\delta (1-\alpha_1-\alpha_2-\cdots -\alpha_{q+1}) \nonumber\\
&\;\;\; \times e^{4\pi i\alpha_1 r_1}e^{4\pi i\alpha_2 r_2}
\cdots e^{4\pi i\alpha_q r_q}e^{4\pi i\alpha_{q+1} r_1}
n^{\frac{q}{2}}(4\pi i)^{q}(-1)^{r_1-j_1+r_2-j_2+\cdots +r_q-j_q} \nonumber\\
&\;\;\; \times C^{R_1m_1}_{j_1r_1j_2-r_2}C^{R_2m_2}_{j_2r_2j_3-r_3}
\cdots C^{R_qm_q}_{j_qr_qj_1-r_1}
{\rm Tr} \langle
H^{(s_1,s_2)}_{R_1m_1}H^{(s_2,s_3)}_{R_2m_2}
\cdots H^{(s_q,s_1)}_{R_qm_q} 
\rangle,
\end{align}
where $C^{Rm}_{jrj'r'}$ is the Clebsch-Gordan coefficient.

\begin{figure}[tbp]
\begin{center}
 \begin{minipage}{0.35\hsize}
  \begin{center}
\includegraphics[width=30mm]{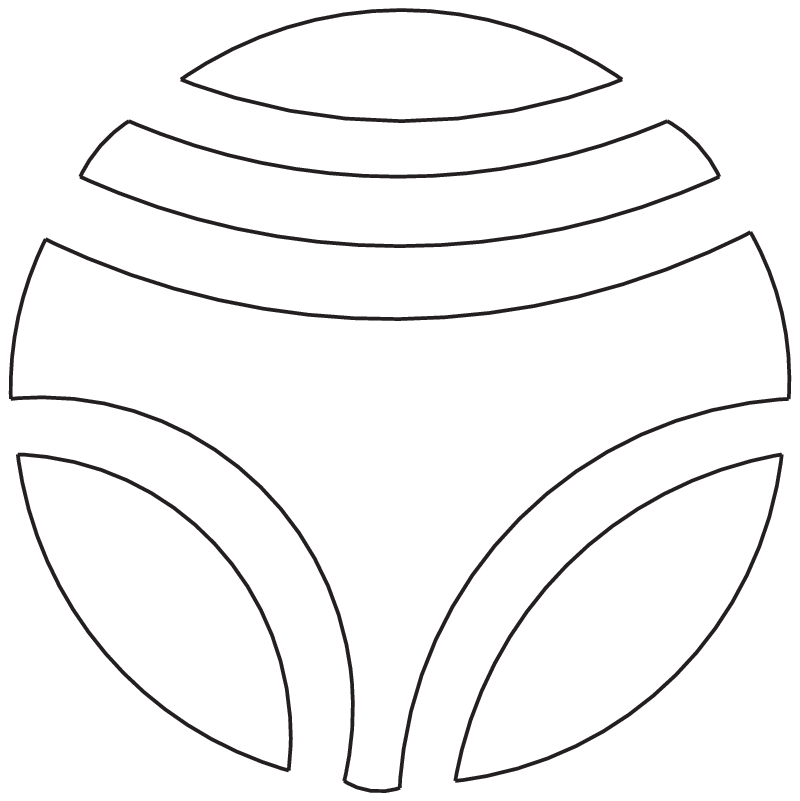}
  \caption{Planar ladder diagram for Wilson loop}
  \label{planar}
\end{center}
 \end{minipage}
\hspace*{1cm}
 \begin{minipage}{0.35\hsize}
  \begin{center}
  {\includegraphics[width=30mm]{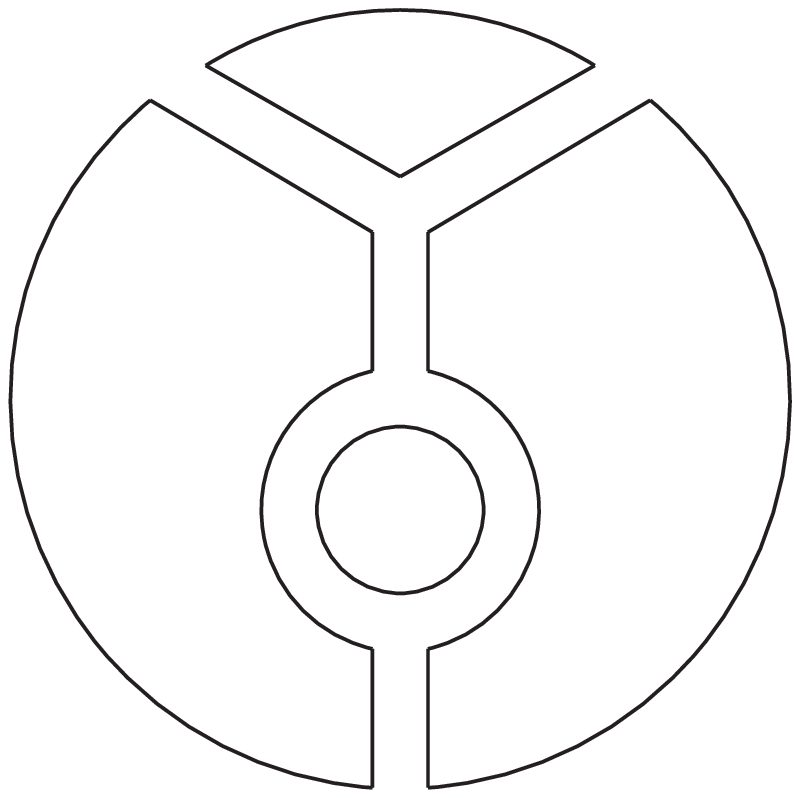}}
  \caption{Planar non-ladder diagram for Wilson loop}
  \label{nonplanar}
\end{center}
 \end{minipage}
\end{center}
\label{Diagrams for Wilson loop}
\end{figure}

In the following, 
we assume that only the planar ladder diagrams (i.e. planar diagrams 
without vertices, see 
Fig. \ref{Diagrams for Wilson loop}) contribute
to the VEV as in the case of the continuum theory. 
This assumption is reasonable since we are working in the gauge 
corresponding to the Feynman gauge on $R^4$.
In the Feynman gauge, the ladder approximation gives an exact result 
in the case of the circular Wilson loop in SYM on $R^4$. 
Hence, also for the PWMM, we believe that the non-ladder 
diagrams do not contribute to the VEV in the continuum limit
in the present gauge.

Within this assumption, only the modes with $R=m=0$ contribute 
to $\langle \hat{W} \rangle$ because the tree-level propagator
(\ref{propagator for H}) is vanishing unless $R=m=0$.
Furthermore, only the terms with $q$ even contribute. 
Thus, we obtain, 
\begin{align}
\langle \hat{W} \rangle 
&=
\frac{1}{\nu k}\sum_{q=0}^{\infty}\sum_{s}
\frac{(4\pi i)^{2q}}{(2q)!}
\left(\frac{n}{2j_s+1}\right)^{q-1}
{\rm Tr}\langle (H_{00}^{(s,s)})^{2q} \rangle,
\end{align}
where we have used 
$C^{00}_{jrj-r}=\frac{1}{\sqrt{2j+1}}$. 
Taking the limit (\ref{continuum limit for N=4}), 
we obtain\footnote{Here, the limit $\nu \rightarrow \infty$ is not
used explicitly. However this limit should be needed for 
the non-ladder diagrams (See Fig.~\ref{nonplanar}) to cancel out.}
\begin{align}
\langle \hat{W} \rangle
&\rightarrow 
\frac{1}{k}
\sum_{q=0}^{\infty}\frac{1}{(2q)!}\langle {\rm Tr} 
(4\pi i H^{(0,0)}_{00})^{2q} \rangle
\nonumber\\
&=\frac{1}{Z}\int {\cal D}M \frac{1}{k} {\rm Tr} \exp(M)
\exp \left(-\frac{k}{\lambda} {\rm Tr} M^2 \right).
\label{abcde}
\end{align}
On the right hand side of the first line of (\ref{abcde}),
we have put $(s,s)$ to $(0,0)$ because the propagator does not
depend on $s$.
In the second line, 
we have put $\lambda=\frac{V_{S^3}g_{PW}^2k}{n}$ and 
identified $4\pi i H_{00}^{(0,0)}$ with a $k\times k$ hermitian matrix $M$.
Although ${H_{00}^{(0,0)}}$ is a complex matrix and time-dependent, 
this identification is possible within the ladder approximation
where only the equal-time propagator in the free theory 
(\ref{propagator for H}) is needed for the computation of the Wilson loop.
Since the equal-time propagator of $4\pi i H_{00}^{(0,0)}$ takes the same 
value as that of $M$ if we take the weight of the integral of $M$
as in (\ref{abcde}), the computational rules are identical between 
the two expressions.
The result (\ref{abcde}) agrees with (\ref{ESZ result}).

Here, we again emphasize that in a general gauge
one needs to add also the planar non-ladder diagrams
(See Fig.~\ref{nonplanar})
to obtain the correct result (\ref{ESZ result}) in the continuum theory. 
In fact, we checked that in a different gauge which will be 
introduced in the next section the sum over only the planar ladder diagrams
does not coincide with (\ref{ESZ result}).
This is consistent with the fact that the value of 
each Feynman diagram depends on a gauge choice and 
the non-ladder diagrams cancel out only in the Feynman gauge on $R^4$.

\section{Beta function}
\setcounter{equation}{0}
In this section, we compute the beta function 
at the 1-loop level in the PWMM expanded around 
the background specified by (\ref{parameters}) and show 
that it is indeed vanishing.
In this section, we take the gauge shown in 
appendix~\ref{Harmonic expansion of PWMM}
for convenience 
which is different from what we used in the previous section.
We consider only the planar diagrams in the following.

\begin{figure}[tbp]
\begin{center}
\includegraphics[width=0.8\textwidth, keepaspectratio, clip]{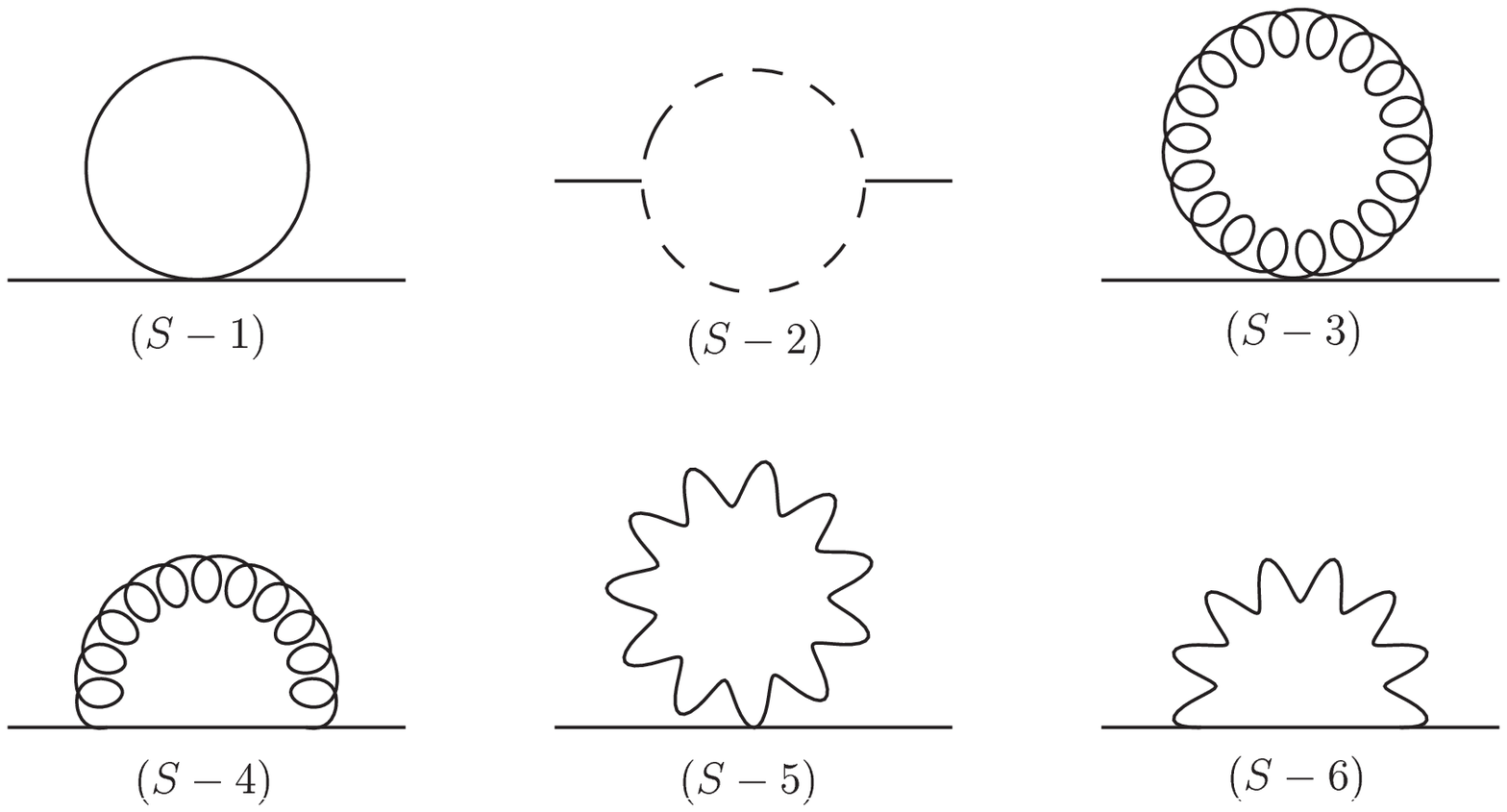}
\end{center}
\caption{The 1-loop corrections to the self-energy of scalar field.
The solid, dashed, curly and wavy lines
represent the propagators of the $SO(6)$ scalar, 
fermion, $SO(3)$ scalar and the gauge field, respectively.}
\label{feynman diagram 1}
\end{figure}

We first compute the wave function renormalization 
of the $SO(6)$ scalar field.
The self-energy is given by the truncated two-point
function $\langle \phi_{ABJm}^{(s,t)}(p)_{ij}
\phi_{CDJ'm'}^{(s',t')}(p')_{kl}  \rangle $ which
takes the form,
\begin{align}
2\pi \delta (p-p')\delta_{st'} \delta_{s't}
\frac{1}{2}\epsilon_{ABCD}\delta_{il}\delta_{jk}
\delta_{JJ'}\delta_{m-m'} \Xi_{J}^{(s,t)}(p).
\label{form of self energy}
\end{align}
There are six diagrams contributing to the 
1-loop correction to the self-energy as shown in 
Fig. \ref{feynman diagram 1}.
For example, one can compute $(S-2)$ in 
Fig. \ref{feynman diagram 1} as follows.
In terms of the Feynman rules shown in 
appendix~\ref{Harmonic expansion of PWMM}, 
the contribution of $(S-2)$ to 
$\Xi_{J}^{(s,t)}(p)$ in (\ref{form of self energy})
can be written as,
\begin{align}
-2k \sum_{J_1m_1\kappa_1}\sum_{J_2m_2\kappa_2}
\int \frac{dq}{2\pi} 
\frac{iq+\kappa_1 \omega_{J_1}^{\psi}}{q^2-(\omega_{J_1}^{\psi})^2}
\frac{i(p-q)+\kappa_2\omega_{J_2}^{\psi}}{(p-q)^2-(\omega_{J_2}{\psi})^2}
\hat{\cal F}^{J_2m_2\kappa_2 (j_uj_t)}_{J_1-m_1\kappa_1(j_uj_s)\; Jm(j_sj_t)}
\hat{\cal F}^{J_1-m_1\kappa_1 (j_uj_s)}_{J_2m_2\kappa_2(j_uj_t)\; Jm(j_tj_s)},
\end{align}
where the 
ranges of the variables in $\sum_{J_im_i\kappa_i}$ 
are given by $\kappa_i=\pm 1$, $m_i=-J_i, -J_i+1, \cdots, J_i$ 
($i=1,2$), $J_1-\kappa_1=|j_s-j_u|,|j_s-j_u|+1,\cdots, j_s+j_u$ 
and $J_2-\kappa_2=|j_t-j_u|,|j_t-j_u|+1,\cdots, j_t+j_u$.
By performing the integration, substituting the
explicit form of $\hat{\cal F}$ shown in (\ref{cF})
and finally taking the summation over 
$m_1,m_2,\kappa_1$ and $\kappa_2$, one can obtain,
\begin{align}
&32\mu  n k (-1)^{m-(j_s-j_t)}
\sum_{R_1=|j_s-j_u|}^{j_s+j_u}
\sum_{R_2=|j_t-j_u|}^{j_t+j_u}
(2J+1)(2R_1+1)(2R_2+1)
\nonumber\\
&\times
\left[ 
\frac{(R_1+1)(R_2+1)(R_1+R_2+\frac{3}{2})}
{p^2+\mu^2(R_1+R_2+\frac{3}{2})^2}
\left\{
\begin{array}{ccc}
R_1+\frac{1}{2} & R_1 & \frac{1}{2}   \\
R_2+\frac{1}{2} & R_2 & \frac{1}{2}   \\
J & J & 0   \\
\end{array}
\right\}
\right.
\nonumber\\
& \left. \hspace{1cm}+
\frac{R_1R_2(R_1+R_2+\frac{1}{2})}
{p^2+\mu^2(R_1+R_2+\frac{1}{2})^2}
\left\{
\begin{array}{ccc}
R_1 &R_1+\frac{1}{2} &  \frac{1}{2}   \\
R_2 &R_2+\frac{1}{2} & \frac{1}{2}   \\
J & J & 0   \\
\end{array}
\right\}
\right]\left\{
\begin{array}{ccc}
R_2 & R_1 & J   \\
j_s & j_t & j_u \\
\end{array}
\right\},
\label{s2}
\end{align}
where $R_i$ are defined as $R_i=J_i-\kappa_i$ for 
$\kappa_i=\pm 1$.
Then, if one sets $J=0$ and $s=t=0$,
it is easy to compute the divergent part of (\ref{s2}). 
In this case, the explicit formula for the 
$6j$-symbol and $9j$-symbol can be found in \cite{vmk}.
Substituting those explicit forms,
one finds that the divergent part is given as 
\begin{align}
\frac{k}{\mu}
\sum_{u=-(\nu-1)/2}^{(\nu-1)/2} \sum_{R=|u/2|}^{n-1+u/2}
\left\{ 8-\frac{2}{R^2} \left( 
\frac{p^2}{\mu^2}+\frac{1}{4} \right) \right\},
\label{S-2}
\end{align}

Similarly, the divergent parts
in the other diagrams in 
Fig. \ref{feynman diagram 1} are evaluated as
\begin{align}
(S-1)&=\frac{k}{\mu}
\sum_{u=-(\nu-1)/2}^{(\nu-1)/2}\sum_{R=|u/2|}^{n-1+u/2}
\left( - 5 \right),
\nonumber\\
(S-3)&=\frac{k}{\mu}
\sum_{u=-(\nu-1)/2}^{(\nu-1)/2}\sum_{R=|u/2|}^{n-1+u/2}
\left( -3+\frac{3}{8R^2} \right),
\nonumber\\
(S-4)&=\frac{k}{\mu}
\sum_{u=-(\nu-1)/2}^{(\nu-1)/2}\sum_{R=|u/2|}^{n-1+u/2}
\left\{ \frac{1}{2} -\frac{1}{8R^2}
\left( \frac{p^2}{\mu^2} +\frac{1}{4} \right) \right\},
\nonumber\\
(S-5)&=\frac{k}{\mu}
\sum_{u=-(\nu-1)/2}^{(\nu-1)/2}\sum_{R=|u/2|}^{n-1+u/2}
\left( -{1} -\frac{1}{8R^2} \right),
\nonumber\\
(S-5)&=\frac{k}{\mu}
\sum_{u=-(\nu-1)/2}^{(\nu-1)/2}\sum_{R=|u/2|}^{n-1+u/2}
\left\{ \frac{1}{2} +\frac{1}{8R^2}
\left( \frac{9p^2}{\mu^2}+\frac{1}{4} \right)  \right\}.
\label{S-other}
\end{align}
Here, it is important that 
the quadratic divergence cancels when 
one takes the sum over all the diagrams
in Fig. \ref{feynman diagram 1}.
Note also that there is no divergence even if one takes 
the large-$n$ limit.
This limit corresponds to the commutative limit
of fuzzy sphere in which the theory on $R\times S^2$
is realized. The fact that there is no divergence
due to $n$ is consistent with the 
super renormalizability of (2+1) dimensional 
gauge theory.

From (\ref{S-2}) and (\ref{S-other}),
we find that the wave function renormalization 
for the $SO(6)$ scalars is given by
\begin{align}
Z_{\phi}=1-\frac{4g^2_{PW} k}{\mu^3 n}\log \nu.
\end{align}
Here, we have introduced the following expression
for the logarithmic divergence,
\begin{align}
\sum_{u=-(\nu-1)/2}^{(\nu-1)/2} \sum_{R=|u/2|}^{\infty}
\frac{1}{R^2} \equiv 4\log \nu.
\end{align}
The wave function renormalization of the 
fermions can be computed in the same way \cite{Ishii:2008ib}.
The result is given by
\begin{align}
Z_{\psi}=1-\frac{16g^2_{PW} k}{\mu^3 n}\log \nu.
\end{align}

\begin{figure}[tbp]
\begin{center}
\includegraphics[width=0.8\textwidth, keepaspectratio, clip]{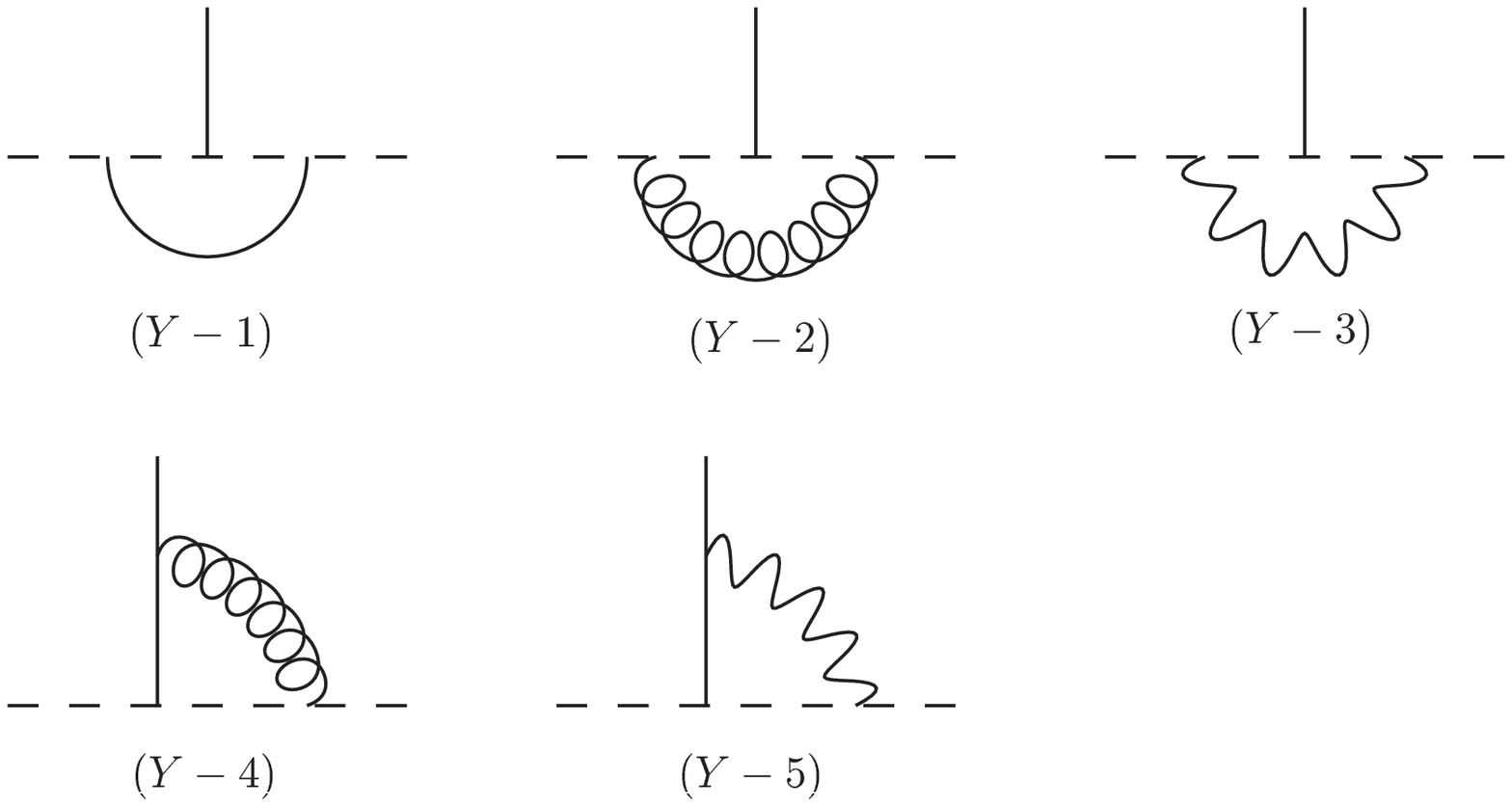}
\end{center}
\caption{The 1-loop corrections to the Yukawa coupling.
The solid, dashed, curly and wavy lines
represent the propagators of the $SO(6)$ scalar, 
fermion, $SO(3)$ scalar and the gauge field, respectively.}
\label{feynman diagram 2}
\end{figure}

Next, we compute the renormalization of the
coupling which can be 
read off from the truncated three point 
function of fermions and a scalar,
$\langle  \phi_{ABJ_1m_1}^{(s_1,t_1)}(p_1) 
\psi^{(s_2,t_2)\dagger }_{C J_2m_2\kappa_2}(p_2)
\psi^{(s_3,t_3)\dagger }_{D J_3m_3\kappa_3}(p_3)
\rangle$.
We take the simplest choice of the external 
momenta: $J_i=0$, $p_i=0$, $s_i=t_i=0$ $(i=1,2,3)$, 
$m_1=0$ and $\kappa_1=\kappa_2=1$.
In Fig. \ref{feynman diagram 2},
all the diagrams contributing to the 1-loop corrections
are listed.
By following the same calculation that we described 
for the self-energy, we find the following 
values for the divergent part of each diagram,
\begin{align}
&(Y-1)=\frac{k}{\mu^3}\epsilon_{ABCD}(-1)^{m+\frac{1}{2}}
\sum_{u=-(\nu-1)/2}^{(\nu-1)/2}\sum_{R=|u/2|}^{n-1+u/2}
\frac{2}{R^2},
\nonumber\\
&(Y-2)=\frac{k}{\mu^3}\epsilon_{ABCD}(-1)^{m+\frac{1}{2}}
\sum_{u=-(\nu-1)/2}^{(\nu-1)/2}\sum_{R=|u/2|}^{n-1+u/2}
\frac{3}{2R^2},
\nonumber\\
&(Y-3)=\frac{k}{\mu^3}\epsilon_{ABCD}(-1)^{m+\frac{1}{2}}
\sum_{u=-(\nu-1)/2}^{(\nu-1)/2}\sum_{R=|u/2|}^{n-1+u/2}
\frac{1}{2R^2},
\nonumber\\
&(Y-4)=\frac{k}{\mu^3}\epsilon_{ABCD}(-1)^{m+\frac{1}{2}}
\sum_{u=-(\nu-1)/2}^{(\nu-1)/2}\sum_{R=|u/2|}^{n-1+u/2}
\frac{3}{4R^2},
\nonumber\\
&(Y-5)=\frac{k}{\mu^3}\epsilon_{ABCD}(-1)^{m+\frac{1}{2}}
\sum_{u=-(\nu-1)/2}^{(\nu-1)/2}\sum_{R=|u/2|}^{n-1+u/2}
\left( -\frac{1}{4R^2}\right) .
\end{align}
Comparing the sum of the above diagrams with the 
tree level result, one obtains the renormalization
for the coupling at the 1-loop level,
\begin{align}
Z_{g_{PW}}=1-\frac{18g^2_{PW} k}{\mu^3n}\log \nu.
\end{align}
This satisfies $Z_{g_{PW}}=Z_{\psi}Z_{\phi}^{\frac{1}{2}}$ and
therefore the beta function is vanishing 
at the 1-loop level.

\section{Conclusion and discussion}
\setcounter{equation}{0}
In this paper, we study a non-perturbative formulation of ${\cal N}=4$ SYM on $R\times S^3$ proposed 
in \cite{Ishii:2008ib}.  
After reviewing the formulation, we introduce Wilson loop operators in the formulation which correspond
to Wilson loops in ${\cal N}=4$ SYM studied in the context of the AdS/CFT correspondence.
We calculate the VEV of the half-BPS Wilson loop by summing up all the planar ladder diagrams
as done in the continuum theory in \cite{Erickson:2000af,Drukker:2000rr}
and reproduce the known result.
We also calculate the one-loop beta function and verify that it vanishes in the continuum limit,
which is consistent with restoration of the superconformal invariance.
Our results serves as a check of the formulation.

We should check that the planar non-ladder diagrams for the half-BPS Wilson loop are indeed canceled out
at lower orders in the perturbative expansion, as done in \cite{Erickson:2000af},
while the calculation in our case seems hard work.
It is also desirable to derive the result using a method such as localization as done in the continuum theory
in \cite{Pestun:2007rz}. 
It is interesting to do the calculation of this paper in another
non-perturbative formulation of ${\cal N}=4$ SYM on $R\times S^3$ proposed in \cite{Kawai:2009vb}.
It is quite important to calculate the VEV of non-BPS Wilson loops such as a rectangular Wilson loop on $R^4$
in the strong coupling regime in the present formulation and 
reproduce the prediction from the gravity side \cite{Maldacena:1998im}.
This kind of calculation on the gauge theory side not done yet would serve as a highly
nontrivial test of the AdS/CFT correspondence. 

We hope that our calculation in this paper will trigger development in the study of
the formulation.

\vspace{0.5cm}

\section*{Note}

Preliminary result on the beta function in this paper
was announced by G. I. and A. T. at the International
Conference on Progress of String Theory and Quantum Fields Theory,
Osaka City University, 7-10 December 2007 \cite{Ishiki,Tsuchiya}.

\section*{Acknowledgements}
We would like to thank T. Ishii for collaboration
at the early stage of this work.
The work of G.I. is supported by the Grant-in-Aid for the Global COE Program ``The Next Generation of Physics, Spun from Universality and Emergence'' from the Ministry of Education, Culture, Sports, Science and Technology (MEXT) of Japan. 
The work of S.S. is supported in part by the JSPS Research Fellowship for Young Scientists.
The work of A.T. is supported in part by Grant-in-Aid for Scientific
Research (No. 19540294) from JSPS.

\appendix

\section{$S^3$ and the $SU(2)$ group manifold}
In this appendix, we summarize some useful facts about $S^3$ and the $SU(2)$ group manifold
(see also \cite{Ishii:2008ib,Ishii:2008tm}).
We regard $S^3$ as the $SU(2)$ group manifold. We parameterize an element
of $SU(2)$ in terms of the Euler angles as
\begin{equation}
g=e^{-i\varphi J_3}e^{-i\theta J_2}e^{-i\psi J_3},
\label{Euler angles}
\end{equation}
where $J_i$ are the generators of the $SU(2)$ algebra, and
$0\leq \theta\leq \pi$, $0\leq \varphi < 2\pi$, $0\leq \psi < 4\pi$.
The periodicity with respect to these angle variables is expressed as
\begin{align}
(\theta,\varphi,\psi)\sim (\theta,\varphi+2\pi,\psi+2\pi)\sim (\theta,\varphi,\psi+4\pi).
\end{align}
The isometry of $S^3$ is $SO(4)=SU(2)\times SU(2)$, and these two
$SU(2)$'s act on $g$ from left and right, respectively. 
We construct the
right-invariant 1-forms, 
\begin{equation}
dgg^{-1}=-i\mu e^i J_i,
\label{right invariant 1-form}
\end{equation}
where the radius of $S^3$ is $2/\mu$. They are explicitly
given by 
\begin{eqnarray}
&&e^1=\frac{1}{\mu}(-\sin \varphi d\theta + \sin\theta\cos\varphi d\psi),\nonumber\\
&&e^2=\frac{1}{\mu}(\cos \varphi d\theta + \sin\theta\sin\varphi
 d\psi),\nonumber\\
&&e^3=\frac{1}{\mu}(d\varphi + \cos\theta d\psi),
\label{explicit form of right invariant 1-form}
\end{eqnarray}
and satisfy the Maurer-Cartan equation
\begin{equation}
de^i-\frac{\mu}{2}\epsilon_{ijk}e^j\wedge e^k=0.\label{Maurer-Cartan}
\end{equation}
The metric is constructed from $e^i$ as
\begin{equation}
ds^2=e^ie^i=\frac{1}{\mu^2}\left(
d\theta^2+\sin^2\theta d\varphi^2 +(d\psi+\cos\theta d\varphi)^2\right).
\label{metric of S^3}
\end{equation}
The Haar measure is defined through above metric as
\begin{align}
d\Omega_3=\frac{1}{8}\sin\theta d\theta d\phi d\psi,
\label{Haar measure}
\end{align}
which is left and right invariant.
The Killing vectors dual to $e^i$ are given by
\begin{equation}
{\cal{L}}_i=-\frac{i}{\mu}e^{\bar{\mu}}_i\partial_{\bar{\mu}},
\label{definition of calL_i}
\end{equation}
where $\bar{\mu}=\theta,\varphi,\psi$ and $e^{\bar{\mu}}_i$ are 
inverse of $e^i_{\bar{\mu}}$.
It follows from (\ref{right invariant 1-form}) and (\ref{definition of calL_i}) that 
\begin{align}
{\cal L}_ig=-J_ig,
\label{left translation}
\end{align}
which indicates that ${\cal L}_i$ are the generators of the left translation 
and satisfy 
\begin{align}
[{\cal{L}}_i,{\cal{L}}_j]=i\epsilon_{ijk}{\cal{L}}_k.
\end{align}
The explicit form of the Killing
vectors are
\begin{eqnarray}
&&{\cal{L}}_1=-i\left(-\sin\varphi\partial_{\theta}-\cot\theta\cos\varphi\partial_{\varphi}+\frac{\cos\varphi}{\sin\theta}\partial_{\psi}\right),\nonumber\\
&&{\cal{L}}_2=-i\left(\cos\varphi\partial_{\theta}-\cot\theta\sin\varphi\partial_{\varphi}+\frac{\sin\varphi}{\sin\theta}\partial_{\psi}\right),\nonumber\\
&&{\cal{L}}_3=-i\partial_{\varphi}.\label{Killing vector}
\end{eqnarray}

The spherical harmonics on $S^3$ is defined through Wigner's D-function (see also \cite{Ishiki:2006rt,Ishiki:2006yr}):
\begin{align}
Y_{Jm\tilde{m}}(\Omega_3)&=(-1)^{J-m}\sqrt{2J+1}D_J(-m,\tilde{m}) \nonumber\\
&=(-1)^{J-m}\sqrt{2J+1}\langle J-m|g|J\tilde{m}\rangle \nonumber\\
&=(-1)^{J-m}\sqrt{2J+1}e^{im\phi-i\tilde{m}\psi}\langle J-m|e^{-i\theta J_2} |J\tilde{m}\rangle,
\label{spherical harmonics on S^3}
\end{align}
where $g$ is given in (\ref{Euler angles}). The complex conjugate is given by
\begin{align}
Y_{Jm\tilde{m}}^*(\Omega_3)
&=(-1)^{J-m}\sqrt{2J+1}e^{-im\phi+i\tilde{m}\psi}\langle J-m|e^{-i\theta J_2}|J\tilde{m}\rangle \nonumber\\
&=(-1)^{J-m}\sqrt{2J+1}e^{-im\phi+i\tilde{m}\psi}(-1)^{-m-\tilde{m}}\langle Jm|e^{-i\theta J_2}|J-\tilde{m}\rangle \nonumber\\
&=(-1)^{m-\tilde{m}}Y_{J-m-\tilde{m}}(\Omega_3).
\end{align}
It follows from (\ref{left translation}) that
\begin{align}
{\cal L}_i^2Y_{Jm\tilde{m}}(\Omega_3)&=J(J+1)Y_{Jm\tilde{m}}(\Omega_3), \nonumber\\
{\cal L}_{\pm}Y_{Jm\tilde{m}}(\Omega_3)&=\sqrt{(J\mp m)(J\pm m+1)}Y_{Jm\pm 1\tilde{m}},\nonumber\\
{\cal L}_3Y_{Jm\tilde{m}}&=mY_{Jm\tilde{m}}.
\end{align}
By using the orthogonality relation
\begin{align}
\int dg \langle J\tilde{m}|g^{-1}|Jm\rangle\langle J'm'|g|J'\tilde{m}'\rangle
=\frac{1}{2J+1}\delta_{JJ'}\delta_{mm'}\delta_{\tilde{m}\tilde{m}},
\end{align}
where $dg=d\Omega_3/(2\pi^2)$ is the Haar measure,
one can show the following equalities:
\begin{align}
\int \frac{d\Omega_3}{2\pi^2} Y_{J_1m_1\tilde{m}_1}^*(\Omega_3)Y_{J_2m2\tilde{m}_2}(\Omega_3)
&=\delta_{J_1J_2}\delta_{m_1m_2}\delta_{\tilde{m}_1\tilde{m}_2}
\label{orthonormality}
\end{align}
and
\begin{align}
{\cal C}^{J_1m_1\tilde{m}_1}_{J_2m_2\tilde{m}_2\,J_3m_3\tilde{m}_3}
&\equiv\int \frac{d\Omega_3}{2\pi^2} Y_{J_1m_1\tilde{m}_1}^*(\Omega_3)Y_{J_2m_2\tilde{m}_2}(\Omega_3)
Y_{J_3m_3\tilde{m}_3}(\Omega_3) \nonumber\\
&=\frac{(2J_2+1)(2J_3+1)}{2J_1+1}C^{J_1m_1}_{J_2m_2J_3m_3}C^{J_1\tilde{m}_1}_{J_2\tilde{m}_2J_3\tilde{m}_3}, \\
{\cal C}_{J_1m_1\tilde{m}_1\,J_2m_2\tilde{m}_2 \,J_3m_3\tilde{m}_3}
&\equiv\int \frac{d\Omega_3}{2\pi^2} Y_{J_1m_1\tilde{m}_1}(\Omega_3)Y_{J_2m_2\tilde{m}_2}(\Omega_3)
Y_{J_3m_3\tilde{m}_3}(\Omega_3) \nonumber\\
&=(-1)^{m_1-\tilde{m}_1}{\cal C}^{J_1-m_1-\tilde{m}_1}_{J_2m_2\tilde{m}_2\,J_3m_3\tilde{m}_3},
\label{calC}
\end{align}
where $C^{J_1m_1}_{J_2m_2J_3m_3}$ is the Clebsch-Gordan coefficient.

\section{Fuzzy spherical harmonics \label{Fuzzy spherical harmonics}}
\setcounter{equation}{0}
In this appendix, 
we summarize some useful properties of the fuzzy spherical harmonics
(See \cite{Ishiki:2006yr,Ishii:2008ib},
and also \cite{Grosse:1995jt,Baez:1998he,de Wit:1988ig,Hoppe:1988gk,Hoppe,Dasgupta:2002hx}).

Given $(2j+1)\times (2j'+1)$ rectangular complex matrices,
we can generally express them as
\begin{align}
M=\sum_{r,r'}M_{rr'}|jr\rangle\langle j'r'|,
\end{align}
where $|jr\rangle$ are the basis of the spin $j$ representation of $SU(2)$ algebra.
$\big\{|jr\rangle\langle j'r'|  \: \big| \: r=-j,\cdots,j ; \: r'=-j',\cdots,j' \big\}$ 
form a basis of $(2j+1)\times (2j'+1)$ matrices.
Let us consider linear maps, which map the set of 
$(2j+1)\times (2j'+1)$ complex matrices into itself, defined by
\begin{align}
L_i \circ |jr\rangle\langle j'r'| \equiv 
L_i^{[j]}|jr\rangle\langle j'r'|-
|jr\rangle\langle j'r'|L_i^{[j']},
\end{align}
where $L_i^{[j]}$ are the spin $j$ representation matrices of
the $SU(2)$ generators. $L_i \circ$ satisfy the $SU(2)$ algebra: 
$[L_i\circ ,L_j\circ] = i\epsilon_{ijk}L_k \circ$.

By changing the basis, we can obtain the more appropriate basis for the action of $L_i \circ$, 
which is called the fuzzy spherical harmonics:
\begin{align}
\hat{Y}_{Jm(jj')}=\sqrt{n}\sum_{r,r'}(-1)^{-j+r'}C^{Jm}_{jr\;j'-r'}
|jr\rangle\langle j'r'|,
\label{fuzzy spherical harmonics}
\end{align}
where $C^{J_1m_1}_{J_2m_2J_3m_3}$ is the Clebsch-Gordan coefficient 
and $n$ is a positive constant, 
which is taken to be an integer as an ultraviolet cutoff in section 2.
For a fixed $J$, the fuzzy spherical harmonics form the basis of
the spin $J$ irreducible representation of $SU(2)$ under $L_i \circ$,
\begin{align}
(L_i \circ)^2\hat{Y}_{Jm(jj')}
&=J(J+1)\hat{Y}_{Jm(jj')},\nonumber\\
L_{\pm}\circ \hat{Y}_{Jm(jj')}
&=\sqrt{(J\mp m)(J \pm
 m+1)}\hat{Y}_{Jm\pm 1(jj')},\nonumber\\
L_3\circ \hat{Y}_{Jm(jj')}
&=m\hat{Y}_{Jm(jj')}.
\label{action of SU(2) on fuzzy spherical harmonics}
\end{align}
The hermitian conjugates of the fuzzy spherical harmonics are given by
\begin{align}
\left(\hat{Y}_{Jm(jj')}\right)^{\dagger}
=(-1)^{m-(j-j')}\hat{Y}_{J-m(j'j)}.
\end{align}
The orthonormality condition is
\begin{align}
\frac{1}{n}{\rm tr}\left\{
\left(\hat{Y}_{Jm(jj')}\right)^{\dagger}
\hat{Y}_{J'm'(jj')}\right\}=\delta_{JJ'}\delta_{mm'},
\label{orthonormality condition for fuzzy spherical harmonics}
\end{align}
where ``${\rm tr}$'' stands for the trace over $(2j'+1)\times(2j'+1)$ matrices.
The trace of three fuzzy spherical harmonics is evaluated as
\begin{align}
\hat{\mathcal{C}}^{J_1m_1(jj'')}_{J_2m_2(jj')J_3m_3(j'j'')}
&\equiv
\frac{1}{n}{\rm tr} \left\{
\left(\hat{Y}_{J_1m_1(jj'')}\right)^{\dagger}
\hat{Y}_{J_2m_2(jj')}\hat{Y}_{J_3m_3(j'j'')}
\right\} \nonumber\\
&=
(-1)^{J_1+j+j''}\sqrt{n(2J_2+1)(2J_3+1)}
C^{J_1m_1}_{J_2m_2J_3m_3}
\left\{
\begin{array}{ccc}
J_1 & J_2 & J_3 \\
j' & j'' & j  
\end{array}
\right\},  \label{Chat1} \\
\hat{\mathcal{C}}_{J_1m_1(j''j)J_2m_2(jj')J_3m_3j'j'')} &\equiv
\frac{1}{n}{\rm tr} \left\{
\hat{Y}_{J_1m_1(j''j)}\hat{Y}_{J_2m_2(jj')}\hat{Y}_{J_3m_3(j'j'')}
\right\} \nonumber\\
&=(-1)^{m_1-(j''-j)}\hat{\mathcal{C}}^{J_1-m_1(jj'')}_{J_2m_2(jj')J_3m_3(j'j'')},
\label{Chat}
\end{align} 
where the last factor of (\ref{Chat1}) is the $6$-$j$ symbol.
The $6$-$j$ symbol is related asymptotically to the Clebsch-Gordan coefficient:
for $R\gg 1$,
\begin{align}
\left\{\begin{array}{ccc}
a & b & c \\
d+R & e+R & f+R 
\end{array} \right\}
\approx \frac{(-1)^{-a+b+2c-d+e+2f+2R}}{\sqrt{2R(2c+1)}}
C^{c \, d-e}_{a\, -e+f\;b\, -f+d}.
\label{6-j symbol and Clebsch-Gordan}
\end{align}

We also introduce the vector fuzzy spherical harmonics $\hat{Y}_{Jm(jj')i}^{\rho}$ 
and the spinor fuzzy spherical harmonics $\hat{Y}_{Jm(jj')\alpha}^{\kappa}$, 
where $\rho$ takes $-1, 0, 1$ and $\kappa$ takes $-1$ and $1$. 
They are defined in terms of the scalar spherical harmonics as
\begin{align}
&\hat{Y}_{Jm(jj')i}^{\rho}
=i^\rho \sum_{n,p} V_{in}\cg{Qm}{\tilde{Q}p}{1n}\hat{Y}_{\tilde{Q}p(jj')}, \n
&\hat{Y}_{Jm(jj')\alpha}^{\kappa}
=\sum_{p} \cg{Um}{\tilde{U}p}{\frac{1}{2}\alpha} \hat{Y}_{\tilde{U}p(jj')},
\label{vector and spinor fuzzy spherical harmonics}
\end{align}
where $Q=J+\delta_{\rho1},\; \tilde{Q}=J+\delta_{\rho-1}$ and 
$U=J+\frac{1}{2}\delta_{\kappa1},\; \tilde{U}=J+\frac{1}{2}\delta_{\kappa-1}$.
$V$ is an unitary matrix given by
\begin{align}
V=\frac{1}{\sqrt{2}}
\begin{pmatrix}
-1 & 0 & 1 \\
-i & 0 & -i \\
0 & \sqrt{2} & 0
\end{pmatrix}.
\end{align}
The vector fuzzy spherical harmonics and the spinor fuzzy spherical harmonics satisfy
\begin{align}
&L_i\circ \hat{Y}_{Jm(jj')i}^{\rho}
=\sqrt{J(J+1)}\delta_{\rho0}\hat{Y}_{Jm(jj')}, \n
&i\epsilon_{ijk}L_j\circ \hat{Y}_{Jm(jj')k}^{\rho}+\hat{Y}_{Jm(jj')i}^{\rho}
=\rho (J+1) \hat{Y}_{Jm(jj')i}^{\rho}, \n
&\left( (\sigma_i)_{\alpha\beta} L_i\circ +\frac{3}{4}\delta_{\alpha\beta}\right)
\hat{Y}_{Jm(jj')\beta}^{\kappa}
=\kappa \left(J+\frac{3}{4}\right) \hat{Y}_{Jm(jj')\alpha}^{\kappa}.
\label{equalities for vector and spinor fuzzy spherical harmonics}
\end{align}
Their hermitian conjugates are given by
\begin{align}
&\left(\hat{Y}_{Jm(jj')i}^{\rho}\right)^\dagger
=(-1)^{m-(j-j')+1}\hat{Y}_{J-m(j'j)i}^{\rho}, \n
&\left(\hat{Y}_{Jm(jj')\alpha}^{\kappa}\right)^\dagger
=(-1)^{m-(j-j')+\kappa \alpha +1}\hat{Y}_{J-m(j'j)-\alpha}^{\kappa},
\end{align}
and the orthonormality conditions are 
\begin{align}
&\frac{1}{n}\tr\left\{
\left(\hat{Y}_{Jm(jj')i}^{\rho}\right)^\dagger
\hat{Y}_{J'm'(jj')i}^{\rho'}\right\}
=\delta_{JJ'}\delta_{mm'}\delta_{\rho\rho'}, \n
&\frac{1}{n}\tr\left\{
\left(\hat{Y}_{Jm(jj')\alpha}^{\kappa}\right)^\dagger
\hat{Y}_{J'm'(jj')\alpha}^{\kappa'}\right\}
=\delta_{JJ'}\delta_{mm'}\delta_{\kappa\kappa'}.
\end{align}
The trace of three fuzzy spherical harmonics, 
including the vector harmonics and/or the spinor harmonics, are evaluated as
\begin{align}
\hat{\cD}^{Jm(j'j)}_{J_1m_1(j'j'')\rho_1\; J_2m_2(j''j)\rho_2}
&\equiv\frac{1}{n}\tr\left\{
\left(\hat{Y}_{Jm(j'j)}\right)^\dagger
\hat{Y}_{J_1m_1(j'j'')i}^{\rho_1}
\hat{Y}_{J_2m_2(j''j)i}^{\rho_2}
\right\} \n
&=\sqrt{3n(2J+1)(2J_1+1)(2J_1+2\rho_1^2+1)(2J_2+1)(2J_2+2\rho_2^2+1)}\n
& \qquad \times (-1)^{\frac{\rho_1+\rho_2}{2}+1+J+j+j'}
\ninej
{Q_1}{\tilde{Q}_1}{1}
{Q_2}{\tilde{Q}_2}{1}
{J}{J}{0}
\cg{Jm}{Q_1m_1}{Q_2m_2}
\sixj
{J}{\tilde{Q}_1}{\tilde{Q}_2}
{j''}{j}{j'}.
\label{cD}
\end{align}
\begin{align}
&\hat{{\cal E}}_{
J_1m_1(jj')\rho_1
J_2m_2(j'j'')\rho_2
J_3m_3(j''j)\rho_3}  \n
&\equiv\epsilon_{ijk}
\frac{1}{n} {\rm tr}
\left(
\hat{Y}^{\rho_1}_{J_1m_1(jj')i}
\hat{Y}^{\rho_2}_{J_2m_2(j'j'')j}
\hat{Y}^{\rho_3}_{J_3m_3(j''j)k}   
\right) \nonumber\\
&=
\sqrt{6n(2J_1+1)(2J_1+2\rho_1^2+1)
       (2J_2+1)(2J_2+2\rho_2^2+1)
       (2J_3+1)(2J_3+2\rho_3^2+1)} \nonumber\\
&\;\;\;
\times (-1)^{-\frac{\rho_1+\rho_2+\rho_3+1}{2}
             -\tilde{Q}_1-\tilde{Q}_2-\tilde{Q}_3
             +2j+2j'+2j''}
\ninej
{Q_1}{\tilde{Q}_1}{1} 
{Q_2}{\tilde{Q}_2}{1}
{Q_3}{\tilde{Q}_3}{1}
\left(
\begin{array}{ccc}
Q_1 & Q_2 & Q_3 \\
m_1 & m_2 & m_3 
\end{array}
\right)
\sixj
{\tilde{Q}_1}{\tilde{Q}_2}{\tilde{Q}_3}
{j''}{j}{j'}.
\label{cE}
\end{align}
\begin{align}
\hat{\cF}^{J_1m_1(j'j)\kappa_1}_{J_2m_2(j'j'')\kappa_2\; Jm(j''j)}
&\equiv
\frac{1}{n}\tr \left\{
\left(\hat{Y}_{J_1m_1(j'j)\alpha}^{\kappa_1}\right)^\dagger
\hat{Y}_{J_2m_2(j'j'')\alpha}^{\kappa_2}
\hat{Y}_{Jm(j''j)}
\right\} \n
&=\sqrt{2n(2\tilde{U}_1+1)(2J+1)^2(2J_2+1)(2J_2+2)}\n
&\qquad \times (-1)^{\tilde{U}_1+j+j'}
\ninej
{U_1}{\tilde{U}_1}{\frac{1}{2}}
{U_2}{\tilde{U}_2}{\frac{1}{2}}
{J}{J}{0}
\cg{U_1m_1}{U_2m_2}{Jm}
\sixj
{\tilde{U}_1}{\tilde{U}_2}{J}
{j''}{j}{j'}.
\label{cF}
\end{align}
\begin{align}
\hat{\cG}^{J_1m_1(j'j)\kappa_1}_{J_2m_2(j'j'')\kappa_2\; Jm(j''j)\rho}
&\equiv
\frac{1}{n}\tr \left\{
\left(\hat{Y}_{J_1m_1(j'j)\alpha}^{\kappa_1}\right)^\dagger
\sigma_{\alpha\beta}^{i}
\hat{Y}_{J_2m_2(j'j'')\beta}^{\kappa_2}
\hat{Y}_{Jm(j''j)i}^{\rho}
\right\} \n
&=\sqrt{6n(2\tilde{U}_1+1)(2J_2+1)(2J_2+2)(2J+1)(2J+2\rho^2+1)}\n
&\qquad \times (-1)^{\frac{\rho}{2}+\tilde{U}_1+j+j'}
\ninej
{U_1}{\tilde{U}_1}{\frac{1}{2}}
{U_2}{\tilde{U}_2}{\frac{1}{2}}
{Q}{\tilde{Q}}{1}
\cg{U_1m_1}{U_2m_2}{Qm}
\sixj
{\tilde{U}_1}{\tilde{U}_2}{\tilde{Q}}
{j''}{j}{j'},
\label{cG}
\end{align}
where $\{\cdots\}$ with $9$ slots is the $9$-$j$ symbol.

\section{Harmonic expansion of PWMM}
\label{Harmonic expansion of PWMM}
\setcounter{equation}{0}

In this appendix, 
we make a harmonic expansion of the PWMM around the vacuum with (\ref{reducible representation}) 
and (\ref{parameters})
for the perturbative analysis in section 4 and 5.

The complete form of the action of the PWMM including fermionic part is given by
\begin{align}
S_{PW}
&=\frac{1}{g_{PW}^2}\int \! d\tau \; \mbox{Tr}
\biggl(
\frac{1}{2}(D_\tau X_i)^2
+\frac{1}{2}\Bigl(X_i+\frac{i}{2}\epsilon_{ijk}[X_j,X_k] \Bigr)^2
+\frac{1}{2}D_\tau \Phi_{AB}D_\tau \Phi^{AB}
+\frac{\mu^2}{8}\Phi_{AB}\Phi^{AB}
\n
&\qquad
-\frac{1}{2}[X_i,\Phi_{AB}][X_i,\Phi^{AB}] 
-\frac{1}{4}[\Phi_{AB},\Phi_{CD}][\Phi^{AB},\Phi^{CD}]
+\psi_A^\dagger D_\tau \psi^A
+\frac{3\mu}{4}\psi_A^\dagger \psi^A \n
&\qquad
+\psi_A^\dagger \sigma^i [X_i,\psi^A]
+\psi_A^\dagger \sigma^2[\Phi^{AB},({\psi}_B^\dagger)^T]
-(\psi^A)^T \sigma^2 [\Phi_{AB},\psi^B] \biggr).
\label{SU(4) PWMM}
\end{align}
where $A,B$ are indices of $\bm{4}$ of $SU(4)$.
$\psi^A$ and $\psi_A^\dagger$ are two-component spinors 
and $\Phi_{AB}$ are $SO(6)$ scalars $X_m$ in (\ref{pp-action}) rewritten
 in terms of $SU(4)$ notation in the following way
\begin{align}
&\Phi_{i4}=\frac{1}{2}(X_{i+3}+iX_{i+6}) \;\;\; (i=1,2,3), \n
&\Phi_{AB}=-\Phi_{BA},\;\;\;\Phi^{AB}=-\Phi^{BA}=\Phi_{AB}^{\dagger},\;\;\;
\Phi^{AB}=\frac{1}{2}\epsilon^{ABCD}\Phi_{CD}. \label{SU(4) and SO(6)}
\end{align}
For this expression, the harmonic expansion can be easily performed.

By replacing $X_i \rightarrow \mu L_i +X_i$ in (\ref{SU(4) PWMM}) 
and adding the gauge fixing and the Fadeev-Popov terms
\begin{align}
S_{gf+gh}=\frac{1}{g_{PW}^2}\int d\tau
\Tr\left\{
\frac{1}{2}\left(\partial_\tau A_\tau+i\mu[L_i,X_i]\right)^2
-i\bar{c} \partial_\tau D_\tau c +\mu \bar{c}[L_i,i\mu[L_i,c]+i[c,X_i]]
\right\}
\end{align}
then we obtain 
\begin{align}
S_{PW+gf+gh}
= S^{gauge}_{PW,free} + S^{gauge}_{PW,int}
+ S^{matter}_{PW,free} + S^{matter}_{PW,int},
\label{gauge fixed action}
\end{align}
where
\begin{align}
S^{gauge}_{PW,free}
&=\frac{1}{g_{PW}^2}\int d\tau \; \mbox{Tr}
\left(
\frac{1}{2}(\partial_\tau X_i)^2-\frac{\mu^2}{2}[L_i,A_\tau]^2
+\frac{1}{2}(\partial_\tau A_\tau)^2
\right. \nonumber\\
&\qquad\qquad\left.
+\frac{\mu^2}{2}(X_i+i\epsilon_{ijk}[L_j,X_k])^2
-\frac{\mu^2}{2}[L_i,X_i]^2
+i\bar{c}\partial_\tau^2 c
-i\mu^2 \bar{c}[L_i,[L_i,c]]
\right),
\label{S_gauge_free}\displaybreak[0] \\
S^{gauge}_{PW,int}
&=\frac{1}{g_{PW}^2}\int d\tau \; \mbox{Tr}
\biggl(
i(\partial_\tau X_i)[A_\tau,X_i]+\mu[A_\tau,X_i][L_i,A_\tau]
+\frac{1}{2}[A_\tau,X_i]^2 \n
&\qquad\qquad
+i\mu\epsilon_{ijk}(X_i+i\epsilon_{ilm}[L_l,X_m])X_j X_k
-\frac{1}{2}\epsilon_{ijk}\epsilon_{ilm}X_j X_k X_l X_m \n
&\qquad \qquad
-i\mu [L_i,\bar{c}][c,X_i]
+\partial_\tau\bar{c}[A_\tau,c]
\biggr),
\label{S_gauge_int}\displaybreak[0]\\
S^{matter}_{PW,free}
&=\frac{1}{g_{PW}^2}\int d\tau \; \mbox{Tr}
\left(
\frac{1}{2}\partial_\tau \Phi_{AB}\partial_\tau \Phi^{AB}
+\frac{\mu^2}{8}\Phi_{AB}\Phi^{AB}
-\frac{\mu^2}{2}[L_i,\Phi_{AB}][L_i,\Phi^{AB}]
\right. \nonumber\\
&\qquad\qquad\left.
+\psi_A^\dagger \partial_\tau \psi^A
+\mu\psi_A^\dagger (\frac{3}{4}\psi^A+\sigma^i [L_i,\psi^A]\big)
\right), \label{S_matter_free}\displaybreak[0]\\
S^{matter}_{PW,int}
&=\frac{1}{g_{PW}^2}\int d\tau \; \mbox{Tr}
\left(
i(\partial_\tau \Phi_{AB})[A_\tau,\Phi^{AB}]
-\frac{1}{2}[A_\tau,\Phi_{AB}][A_\tau,\Phi^{AB}]
-\mu[L_i,\Phi_{AB}][X_i,\Phi^{AB}]
\right. \nonumber\\
&\qquad\qquad
-\frac{1}{2}[X_i,\Phi_{AB}][X_i,\Phi^{AB}] 
-\frac{1}{4}[\Phi_{AB},\Phi_{CD}][\Phi^{AB},\Phi^{CD}]
+i \psi_A^\dagger [A_\tau,\psi^A]
\nonumber\\
&\qquad\qquad\left.
+\psi_A^\dagger \sigma^i [X_i,\psi^A]
+\psi_A^\dagger \sigma^2[\Phi^{AB},({\psi}_B^\dagger)^T]
-(\psi^A)^T \sigma^2 [\Phi_{AB},\psi^B]
\right).\label{S_matter_int}
\end{align}
We perform a mode expansion for each $(s,t)$ block for each field 
in terms of the fuzzy spherical harmonics defined in appendix A:
\begin{align}
A_\tau^{(s,t)}
&=\sum_{J=|j_s-j_t|}^{j_s+j_t}\sum_{m=-J}^{J}B^{(s,t)}_{Jm}
\otimes\hat{Y}_{Jm(j_sj_t)}, 
\qquad
\Phi_{AB}^{(s,t)}
=\sum_{J=|j_s-j_t|}^{j_s+j_t}\sum_{m=-J}^{J}\phi^{(s,t)}_{AB,Jm}\otimes
\hat{Y}_{Jm(j_sj_t)},
\nonumber\displaybreak[0]\\
c^{(s,t)}
&=\sum_{J=|j_s-j_t|}^{j_s+j_t}\sum_{m=-J}^{J}c^{(s,t)}_{Jm}
\otimes\hat{Y}_{Jm(j_sj_t)}, 
\qquad
\bar{c}^{(s,t)}
=\sum_{J=|j_s-j_t|}^{j_s+j_t}\sum_{m=-J}^{J}\bar{c}^{(s,t)}_{Jm}\otimes
\hat{Y}_{Jm(j_sj_t)},
\nonumber\displaybreak[0]\\
\psi^{A(s,t)}
&=\sum_{\kappa=\pm1}\sum_{\tilde{U}=|j_s-j_t|}^{j_s+j_t}\sum_{m=-U}^{U}
\psi_{Jm\kappa}^{A(s,t)}\otimes\hat{Y}_{Jm(j_sj_t)}^\kappa \nonumber\\
&=\sum_{J=|j_s-j_t|}^{j_s+j_t}\sum_{m=-J-\frac{1}{2}}^{J+\frac{1}{2}}
\psi_{Jm1}^{A(s,t)}\otimes\hat{Y}_{Jm(j_sj_t)}^1
+\sum_{J=|j_s-j_t|-\frac{1}{2}}^{j_s+j_t-\frac{1}{2}}\sum_{m=-J}^{J}
\psi_{Jm-1}^{A(s,t)}\otimes\hat{Y}_{Jm(j_sj_t)}^{-1},
\nonumber\displaybreak[0]\\
\psi_A^{(t,s)\dagger}
&=\sum_{\kappa=\pm1}\sum_{\tilde{U}=|j_s-j_t|}^{j_s+j_t}\sum_{m=-U}^{U}
\psi_{A,Jm\kappa}^{(t,s)\dagger}\otimes\hat{Y}_{Jm(j_tj_s)}^{\kappa\dagger} \nonumber\\
&=\sum_{J=|j_s-j_t|}^{j_s+j_t}\sum_{m=-J-\frac{1}{2}}^{J+\frac{1}{2}}
\psi_{A,Jm1}^{(t,s)\dagger}\otimes\hat{Y}_{Jm(j_tj_s)}^{1\dagger}
+\sum_{J=|j_s-j_t|-\frac{1}{2}}^{j_s+j_t-\frac{1}{2}}\sum_{m=-J}^{J}
\psi_{A,Jm-1}^{(t,s)\dagger}\otimes\hat{Y}_{Jm(j_tj_s)}^{-1\dagger},
\nonumber\displaybreak[0]\\
X_i^{(s,t)}
&=\sum_{\rho=-1}^{1}\sum_{\tilde{Q}=|j_s-j_t|}^{j_s+j_t}\sum_{m=-Q}^{Q}
x_{Jm\rho}^{(s,t)}\otimes\hat{Y}{}_{Jm(j_sj_t)i}^\rho \nonumber\\
&=\sum_{J=|j_s-j_t|}^{j_s+j_t}\sum_{m=-J-1}^{J+1}
x_{Jm1}^{(s,t)}\otimes\hat{Y}{}_{Jm(j_sj_t)i}^1 
+\sum_{J=|j_s-j_t|}^{j_s+j_t}\sum_{m=-J}^{J}
x_{Jm0}^{(s,t)}\otimes\hat{Y}{}_{Jm(j_sj_t)i}^0 \nonumber\\
&\;\;\;\;\;+\sum_{J=|j_s-j_t|-1}^{j_s+j_t-1}\sum_{m=-J}^{J}
x_{Jm-1}^{(s,t)}\otimes\hat{Y}{}_{Jm(j_sj_t)i}^{-1}.
\label{mode_expansion_in_PWMM}
\end{align}
Note that the modes in the right-hand sides are $(2j_s+1) \times (2j_t+1)$ matrices.
The free part of (\ref{gauge fixed action}) is expressed in terms of the modes as:
\begin{align}
&S^{gauge}_{PW,free}+S^{matter}_{PW,free}\nonumber\\
&=\frac{n}{g_{PW}^2} \int d\tau \; \mbox{tr}
\Biggl(
\frac{1}{2}(-1)^{m-(j_s-j_t)+1}x_{Jm\rho}^{(s,t)}
\bigl\{-\partial_\tau^2+\rho^2{\omega_J^x}^2+\mu^2\delta_{\rho0}J(J+1)\bigr\}
x_{J-m\rho}^{(t,s)} \n
&\qquad\qquad
+\frac{1}{2}(-1)^{m-(j_s-j_t)}B_{Jm}^{(s,t)}
\bigl\{-\partial_\tau^2+\mu^2 J(J+1)\bigr\}B_{J-m}^{(t,s)}\n
&\qquad\qquad
+i(-1)^{m-(j_s-j_t)}\bar{c}_{Jm}^{(s,t)}
\bigl\{\partial_\tau^2-\mu^2 J(J+1)\bigr\}c_{J-m}^{(t,s)}\n
&\qquad\qquad
+\frac{1}{4}(-1)^{m-(j_s-j_t)}\epsilon^{ABCD}\phi_{AB,Jm}^{(s,t)}
(-\partial_\tau^2+{\omega_J^x}^2)
\phi_{CD,J-m}^{(t,s)}
+\psi_{A,Jm\kappa}^{(s,t)\dagger}
(\partial_\tau+\kappa\omega_J^\psi)\psi_{Jm\kappa}^{A(s,t)}
\Biggr),
\label{harmonic_expansion_of_S_free}
\end{align}
where
\begin{align}
 \omega_J^x\equiv\mu(J+1),\quad
 \omega_J^\psi\equiv\mu(J+\frac{3}{4}),\quad
\omega_J^\phi\equiv\mu(J+\frac{1}{2}).
\label{mass}
\end{align}
From (\ref{harmonic_expansion_of_S_free}), the propagators can be read off as
\begin{align}
&\langle x_{Jm\rho}^{(s,t)}(p)_{ij} x_{J'm'\rho'}^{(s',t')}(p')_{kl}\rangle
\n
&=
\begin{cases}
\frac{g_{PW}^2}{n}(-1)^{m-(j_s-j_t)+1}\delta_{JJ'}\delta_{m\,-m'}\delta_{\rho\rho'}
\delta_{st'}\delta_{ts'}\delta_{il}\delta_{jk}
2\pi\delta(p+p')\frac{1}{p^2+{\omega_J^x}^2}
\;\;(\rho\neq0)\\
\frac{g_{PW}^2}{n} (-1)^{m-(j_s-j_t)+1}\delta_{JJ'}\delta_{m\,-m'}
\delta_{st'}\delta_{ts'}\delta_{il}\delta_{jk}
2\pi\delta(p+p')\frac{1}{p^2+\mu^2J(J+1)}
\;\;(\rho=\rho'=0 )
\end{cases},
\displaybreak[0]\n
&\langle B_{Jm}^{(s,t)}(p)_{ij} B_{J'm'}^{(s',t')}(p')_{kl}\rangle
=
\frac{g_{PW}^2}{n}(-1)^{m-(j_s-j_t)}\delta_{JJ'}\delta_{m\,-m'}
\delta_{st'}\delta_{ts'}\delta_{il}\delta_{jk}
2\pi\delta(p+p')\frac{1}{p^2+\mu^2 J(J+1)},
\displaybreak[0]\n
&\langle c_{Jm}^{(s,t)}(p)_{ij} \bar{c}_{J'm'}^{(s',t')}(p')_{kl}\rangle
=
\frac{g_{PW}^2}{n}(-1)^{m-(j_s-j_t)}\delta_{JJ'}\delta_{m\,-m'}
\delta_{st'}\delta_{ts'}\delta_{il}\delta_{jk}
2\pi\delta(p+p')\frac{i}{p^2+\mu^2 J(J+1)},
\displaybreak[0]\n
&\langle \phi_{AB,Jm}^{(s,t)}(p)_{ij} \phi_{A'B',J'm'}^{(s',t')}(p')_{kl}
\rangle
=\frac{g_{PW}^2}{n} \frac{1}{2}\epsilon_{ABA'B'}
(-1)^{m-(j_s-j_t)}\delta_{JJ'}\delta_{m\,-m'}
\delta_{st'}\delta_{ts'}\delta_{il}\delta_{jk}
2\pi\delta(p+p')\frac{1}{p^2+{\omega_J^\phi}^2},
\displaybreak[0]\n
&\langle \psi_{Jm\kappa}^{A(s,t)}(p)_{ij} \psi_{A',J'm'\kappa'}^{(s',t')\dagger}(p')_{kl}\rangle
=\frac{g_{PW}^2}{n} \delta_{JJ'}\delta_{mm'}\delta_{\kappa\kappa'}\delta^A_{A'}
\delta_{ss'}\delta_{tt'}\delta_{il}\delta_{jk}
2\pi\delta(p-p')\frac{(ip+\kappa\omega_J^\psi)}{p^2+{\omega_J^\psi}^2}.
\label{propagators}
\end{align}

The gauge interaction terms in (\ref{gauge fixed action}) are rewritten as
\begin{align}
&S^{gauge}_{PW,int}\nonumber\\
&=\frac{n}{g_{PW}^2} \int d\tau \; \mbox{tr}
\Big[
i\hat{\mathcal{D}}_{J_1m_1(j_sj_t)\;J_2m_2(j_tj_u)\rho_2\;J_3m_3(j_uj_s)\rho_3}
B_{J_1m_1}^{(s,t)}
(\partial_\tau x_{J_2m_2\rho_2}^{(t,u)} x_{J_3m_3\rho_3}^{(u,s)}
-x_{J_2m_2\rho_2}^{(t,u)} \partial_\tau x_{J_3m_3\rho_3}^{(u,s)})
\nonumber \displaybreak[0]\\
&-\mu
(\sqrt{J_2(J_2+1)}\hat{\mathcal{D}}_{J_1m_1(j_sj_t)\;J_2m_2(j_tj_u)0\;J_3m_3(j_uj_s)\rho_3}
-\sqrt{J_1(J_1+1)}\hat{\mathcal{D}}_{J_2m_2(j_tj_u)\;J_3m_3(j_uj_s)\rho_3\;J_1m_1(j_sj_t)0})
\nonumber\\
&\quad\times
B_{J_1m_1}^{(s,t)} B_{J_2m_2}^{(t,u)} x_{J_3m_3\rho_3}^{(u,s)}
\nonumber\displaybreak[0]\\
&+(-1)^{m-(j_s-j_u)+1}
\nonumber\\
&\quad\times(\hat{\mathcal{D}}_{J_1m_1(j_sj_t)\;J_2m_2(j_tj_u)\rho_2\;J-m(j_uj_s)\rho}
\hat{\mathcal{D}}_{J_4m_4(j_vj_s)\;Jm(j_sj_t)\rho\;J_3m_3(j_uj_v)\rho_3}
B_{J_1m_1}^{(s,t)}  x_{J_2m_2\rho_2}^{(t,u)} x_{J_3m_3\rho_3}^{(u,v)}   B_{J_4m_4}^{(v,s)}
\nonumber\\
&\quad-\hat{\mathcal{D}}_{J_1m_1(j_sj_t)\;J_2m_2(j_tj_u)\rho_2\;J-m(j_uj_s)\rho}
\hat{\mathcal{D}}_{J_3m_3(j_uj_v)\;J_4m_4(j_vj_s)\rho_4\;Jm(j_sj_u)\rho}
B_{J_1m_1}^{(s,t)}  x_{J_2m_2\rho_2}^{(t,u)} B_{J_3m_3}^{(u,v)} x_{J_4m_4\rho_4}^{(v,s)})
\nonumber\displaybreak[0]\\
&+i\mu\rho_1(J_1+1)
\hat{\mathcal{E}}_{J_1m_1(j_sj_t)\rho_1\;J_2m_2(j_tj_u)\rho_2\;J_3m_3(j_uj_s)\rho_3}
x_{J_1m_1\rho_1}^{(s,t)} x_{J_2m_2\rho_2}^{(t,u)} x_{J_3m_3\rho_3}^{(u,s)}
\nonumber\displaybreak[0]\\
&-\frac{1}{2}(-1)^{m-(j_s-j_u)+1}
\hat{\mathcal{E}}_{J-m(j_uj_s)\rho\;J_1m_1(j_sj_t)\rho_1\;J_2m_2(j_tj_u)\rho_2}
\hat{\mathcal{E}}_{Jm(j_sj_u)\rho\;J_3m_3(j_uj_v)\rho_3\;J_4m_4(j_vj_s)\rho_4}
\nonumber\\
&\quad\times
x_{J_1m_1\rho_1}^{(s,t)} x_{J_2m_2\rho_2}^{(t,u)}
x_{J_3m_3\rho_3}^{(u,v)} x_{J_4m_4\rho_4}^{(v,s)}
\nonumber\displaybreak[0]\\
&-i\mu\sqrt{J_3(J_3+1)}\;
(\hat{\mathcal{D}}_{J_2m_2(j_sj_t)\;J_3m_3(j_tj_u)0\;J_1m_1(j_uj_s)\rho_1}
c_{J_2m_2}^{(s,t)} \bar{c}_{J_3m_3}^{(t,u)} x_{J_1m_1\rho_1}^{(u,s)}
\nonumber\\
&\quad-\hat{\mathcal{D}}_{J_2m_2(j_sj_t)\;J_1m_1(j_tj_u)\rho_1\;J_3m_3(j_uj_s)0}
c_{J_2m_2}^{(s,t)} x_{J_1m_1\rho_1}^{(t,u)} \bar{c}_{J_3m_3}^{(u,s)})
\nonumber\displaybreak[0]\\
&-\hat{\mathcal{C}}_{J_1m_1(j_sj_t)\;J_2m_2(j_tj_u)\;J_3m_3(j_uj_s)}
B_{J_1m_1}^{(s,t)}
(\partial_\tau \bar{c}_{J_2m_2}^{(t,u)} c_{J_3m_3}^{(u,s)}
+c_{J_2m_2}^{(t,u)}\partial_\tau \bar{c}_{J_3m_3}^{(u,s)})
\Big].
\label{harmonic_expansion_of_S_gauge_int}
\end{align}
The matter interaction terms in (\ref{gauge fixed action}) are rewritten as
{\small
\begin{align}
&S^{matter}_{PW,int}\nonumber\\
&=\frac{n}{g_{PW}^2} \int d\tau \; \mbox{tr}
\bigg[-\frac{i}{2}\epsilon^{ABCD}
\hat{\mathcal{C}}_{J_1m_1(j_sj_t)\;J_2m_2(j_tj_u)\;J_3m_3(j_uj_s)}
B_{J_1m_1}^{(s,t)}
(\partial_\tau \phi_{AB,J_2m_2}^{(t,u)} \phi_{CD,J_3m_3}^{(u,s)}
-\phi_{AB,J_2m_2}^{(t,u)}\partial_\tau \phi_{CD,J_3m_3}^{(u,s)})
\nonumber \displaybreak[0]\\
&+\frac{1}{2}\epsilon^{ABCD}
\hat{\mathcal{C}}^{Jm(j_sj_u)}_{J_1m_1(j_sj_t)\;J_2m_2(j_tj_u)}
\hat{\mathcal{C}}_{Jm(j_sj_u)\;J_3m_3(j_uj_v)\;J_4m_4(j_vj_s)}
\nonumber\\
&\quad\times
(B_{J_1m_1}^{(s,t)} B_{J_2m_2}^{(t,u)} \phi_{AB,J_3m_3}^{(u,v)} \phi_{CD,J_4m_4}^{(v,s)}
-B_{J_1m_1}^{(s,t)} \phi_{AB,J_2m_2}^{(t,u)} B_{J_3m_3}^{(u,v)} \phi_{CD,J_4m_4}^{(v,s)})
\nonumber\displaybreak[0]\\
&-\frac{\mu}{2}\epsilon^{ABCD}
(\sqrt{J_2(J_2+1)}
\hat{\mathcal{D}}_{J_1m_1(j_sj_t)\;J_2m_2(j_tj_u)0\;J_3m_3(j_uj_s)\rho_3}
\nonumber\\
&\quad-\sqrt{J_1(J_1+1)}
\hat{\mathcal{D}}_{J_2m_2(j_tj_u)\;J_3m_3(j_uj_s)\rho_3\;J_1m_1(j_sj_t)0})
 \phi_{AB,J_1m_1}^{(s,t)} \phi_{CD,J_2m_2}^{(t,u)} x_{J_3m_3\rho_3}^{(u,s)}
\nonumber\displaybreak[0]\\
&-\frac{1}{2}\epsilon^{ABCD}(-1)^{m-(j_s-j_u)+1}
\nonumber\\
&\quad\times
(\hat{\mathcal{D}}_{J_4m_4(j_vj_s)\;Jm(j_sj_u)\rho\;J_3m_3(j_uj_v)\rho_3}
\hat{\mathcal{D}}_{J_2m_2(j_tj_u)\;J-m(j_uj_s)\rho\;J_1m_1(j_sj_t)\rho_1}
 x_{J_1m_1\rho_1}^{(s,t)} \phi_{AB,J_2m_2}^{(t,u)}
 x_{J_3m_3\rho_3}^{(u,v)} \phi_{CD,J_4m_4}^{(v,s)}
\nonumber\\
&\quad
-\hat{\mathcal{D}}_{J_4m_4(j_uj_v)\;J_3m_3(j_vj_s)\rho_3\;Jm(j_sj_u)\rho}
\hat{\mathcal{D}}_{J_2m_2(j_tj_u)\;J-m(j_uj_s)\rho\;J_1m_1(j_sj_t)\rho_1}
 x_{J_1m_1\rho_1}^{(s,t)} \phi_{AB,J_2m_2}^{(t,u)}
 \phi_{CD,J_4m_4}^{(u,v)} x_{J_3m_3\rho_3}^{(v,s)})
\nonumber\displaybreak[0]\\
&-\frac{1}{8}\epsilon^{ABEF}\epsilon^{CDGH}
\hat{\mathcal{C}}^{Jm(j_sj_u)}_{J_1m_1(j_sj_t)\;J_2m_2(j_tj_u)}
\hat{\mathcal{C}}_{Jm(j_sj_u)\;J_3m_3(j_uj_v)\;J_4m_4(j_vj_s)}
\nonumber\\
&\quad\times
(\phi_{AB,J_1m_1}^{(s,t)} \phi_{CD,J_2m_2}^{(t,u)} \phi_{EF,J_3m_3}^{(u,v)} \phi_{GH,J_4m_4}^{(v,s)}
-\phi_{AB,J_1m_1}^{(s,t)} \phi_{EF,J_2m_2}^{(t,u)} \phi_{CD,J_3m_3}^{(u,v)} \phi_{GH,J_4m_4}^{(v,s)})
\nonumber\displaybreak[0]\\
&+i\left((-1)^{m_3-(j_s-j_u)+\frac{\kappa_1-\kappa_2}{2}}
\hat{\mathcal{F}}^{J_2-m_2(j_tj_u)\kappa_2}_{J_1-m_1(j_tj_s)\kappa_1\;J_3m_3(j_sj_u)}
\psi_{A,J_1m_1\kappa_1}^{(s,t)\dagger} B_{J_3m_3}^{(s,u)} \psi_{J_2m_2\kappa_2}^{A(u,t)}
\right.\nonumber\\
&\quad\left.
-\hat{\mathcal{F}}^{J_1m_1(j_sj_t)\kappa_1}_{J_2m_2(j_sj_u)\kappa_2\;J_3m_3(j_uj_t)}
\psi_{A,J_1m_1\kappa_1}^{(s,t)\dagger} \psi_{J_2m_2\kappa_2}^{A(s,u)} B_{J_3m_3}^{(u,t)}
\right)\nonumber\displaybreak[0]\\
&-\left((-1)^{m_3-(j_s-j_u)+\frac{\kappa_1-\kappa_2}{2}}
\hat{\mathcal{G}}^{J_2-m_2(j_tj_u)\kappa_2}_{J_1-m_1(j_tj_s)\kappa_1\;J_3m_3(j_sj_u)\rho_3}
\psi_{A,J_1m_1\kappa_1}^{(s,t)\dagger} x_{J_3m_3\rho_3}^{(s,u)} \psi_{J_2m_2\kappa_2}^{A(u,t)}
\right.\nonumber\\
&\quad\left.
+\hat{\mathcal{G}}^{J_1m_1(j_sj_t)\kappa_1}_{J_2m_2(j_sj_u)\kappa_2\;J_3m_3(j_uj_t)\rho_3}
\psi_{A,J_1m_1\kappa_1}^{(s,t)\dagger} \psi_{J_2m_2\kappa_2}^{A(s,u)} x_{J_3m_3\rho_3}^{(u,t)}
\right)\nonumber\displaybreak[0]\\
&-\frac{i}{2}\epsilon^{ABCD}
\left(
(-1)^{m_1-(j_s-j_t)-\frac{\kappa_1}{2}}
\hat{\mathcal{F}}^{J_2m_2(j_tj_u)\kappa_2}_{J_1-m_1(j_tj_s)\kappa_1\;J_3m_3(j_sj_u)}
\psi_{A,J_1m_1\kappa_1}^{(s,t)\dagger} \phi_{CD,J_3m_3}^{(s,u)} \psi_{B,J_2m_2\kappa_2}^{(t,u)\dagger}
\right.\nonumber\\
&\quad\left.
+(-1)^{m_2-(j_u-j_s)-\frac{\kappa_2}{2}}
\hat{\mathcal{F}}^{J_1m_1(j_sj_t)\kappa_1}_{J_2-m_2(j_sj_u)\kappa_2\;J_3m_3(j_uj_t)}
\psi_{A,J_1m_1\kappa_1}^{(s,t)\dagger} \psi_{B,J_2m_2\kappa_2}^{(u,s)\dagger}
\phi_{CD,J_3m_3}^{(u,t)}
\right)\nonumber\displaybreak[0]\\
&-i\left((-1)^{m_2-(j_u-j_s)-\frac{\kappa_2}{2}}
\hat{\mathcal{F}}^{J_2-m_2(j_sj_u)\kappa_2}_{J_1m_1(j_sj_t)\kappa_1\;J_3m_3(j_tj_u)}
\psi_{J_1m_1\kappa_1}^{A,(s,t)} \phi_{AB,J_3m_3}^{(t,u)} \psi_{J_2m_2\kappa_2}^{B(u,s)}
\right.\nonumber\\
&\quad\left.
+(-1)^{m_1-(j_s-j_t)-\frac{\kappa_1}{2}}
\hat{\mathcal{F}}^{J_1-m_1(j_tj_s)\kappa_1}_{J_2m_2(j_tj_u)\kappa_2\;J_3m_3(j_uj_s)}
\psi_{J_1m_1\kappa_1}^{A(s,t)} \psi_{J_2m_2\kappa_2}^{B(t,u)} \phi_{AB,J_3m_3}^{(u,s)}
\right)
\bigg].
\label{harmonic_expansion_of_S_matter_int}
\end{align}
}

\section{Generalization of the Feynman formula}
\label{Generalization of the Feynman formula}
\setcounter{equation}{0}
We first show that for $p_1,\cdots,p_{k}> 0$
\begin{align}
\int_0^1d\tau_1\cdots\int_0^1d\tau_{k}\delta(1-\tau_1-\cdots-\tau_{k})\tau_1^{p_1-1}\cdots\tau_{k}^{p_k-1}
=\frac{\Gamma(p_1)\cdots\Gamma(p_k)}{\Gamma(p_1+\cdots+p_k)}.
\label{gammafn}
\end{align}
The left-hand side of (\ref{gammafn}) equals
\begin{align}
\int_0^1d\tau_1\int_0^{1-\tau_1}d\tau_2\cdots\int_0^{1-\tau_1-\cdots-\tau_{k-2}}d\tau_{k-1}
\tau_1^{p_1-1}\cdots\tau_{k-1}^{p_{k-1}-1}(1-\tau_1-\cdots-\tau_{k-1})^{p_k-1}.
\end{align}
By putting $\tau_{k-1}=(1-\tau_1-\cdots-\tau_{k-2})\sigma_{k-1}$, we calculate this as 
\begin{align}
\beta(p_{k-1},p_k)\int_0^1d\tau_1\int_0^{1-\tau_1}d\tau_2\cdots\int_0^{1-\tau_1-\cdots-\tau_{k-3}}d\tau_{k-2}
\tau_1^{p_1-1}\cdots\tau_{k-2}^{p_{k-2}-1}(1-\tau_1-\cdots-\tau_{k-2})^{p_{k-1}+p_k-1}.
\end{align}
By repeating this procedure, we see that the left-hand side of (\ref{gammafn}) equals
\begin{align}
\beta(p_{k-1},p_k)\beta(p_{k-2},p_{k-1}+p_k)\cdots\beta(p_1,p_2+\cdots+p_k).
\end{align}
By using 
\begin{align}
\beta(a,b)=\frac{\Gamma(a)\Gamma(b)}{\Gamma(a+b)},
\end{align}
we see that this is indeed equal to the right-hand side of (\ref{gammafn}).

Here we would like to show that
\begin{align}
&\left.\frac{1}{m!}\frac{d^m}{dg^m}e^{A+gB}\right|_{g=0} \nonumber\\
&=\int_0^1d\alpha_1\cdots\int_0^1d\alpha_{m+1}\delta(1-\alpha_1-\cdots-\alpha_{m+1})
e^{\alpha_1A}Be^{\alpha_2A}B\cdots Be^{\alpha_mA}Be^{\alpha_{m+1}A} \nonumber\\
&=\int_0^1d\alpha_1\int_0^{1-\alpha_1}d\alpha_2\cdots\int_0^{1-\alpha_1-\cdots-\alpha_{m-1}}
d\alpha_m
e^{\alpha_1A}Be^{\alpha_2A}B\cdots Be^{\alpha_mA}Be^{(1-\alpha_1-\cdots-\alpha_m)A}.
\label{formula}
\end{align}
By expanding the exponential and using (\ref{gammafn}), we calculate the second line of (\ref{formula}) as
\begin{align}
&\sum_{l_1,\cdots,l_{m+1}=0}^{\infty}\frac{1}{l_1!\cdots l_{m+1}!}
\int_0^1d\alpha_1\cdots\int_0^1d\alpha_{m+1}\delta(1-\alpha_1-\cdots-\alpha_{m+1})
\alpha_1^{l_1}\cdots\alpha_{m+1}^{l_{m+1}}  \nonumber\\
&\qquad\qquad\qquad\qquad\qquad\qquad \times A^{l_1}BA^{l_2}B\cdots BA^{l_m}BA^{l_{m+1}} \nonumber\\
&=\sum_{l_1,\cdots,l_{m+1}=0}^{\infty}\frac{1}{l_1!\cdots l_{m+1}!}
\frac{\Gamma(l_1+1)\cdots\Gamma(l_{m+1}+1)}{\Gamma(l_1+\cdots+l_{m+1}+m+1)}
A^{l_1}BA^{l_2}B\cdots BA^{l_m}BA^{l_{m+1}}  \nonumber\\
&=\sum_{l_1,\cdots,l_{m+1}=0}^{\infty}\frac{1}{(l_1+\cdots+l_{m+1}+m)!}A^{l_1}BA^{l_2}B\cdots BA^{l_m}BA^{l_{m+1}} 
\nonumber\\
&=\sum_{n=m}^{\infty}\frac{1}{n!}\sum_{l_1+\cdots+l_{m+1}=n-m,\;l_1,\cdots,l_{m+1}\geq 0}
A^{l_1}BA^{l_2}B\cdots BA^{l_m}BA^{l_{m+1}}.
\end{align}
This agrees with the left-hand side of (\ref{formula}).

\end{document}